\documentclass{aa}  

\usepackage{graphicx}
\usepackage{txfonts}
\begin{document}

   \title{Lithium abundance patterns of late-F stars: \\an in-depth analysis of the lithium desert \thanks{Tables 1 and 3 are only available in electronic form at the CDS via anonymous ftp to cdsarc.u-strasbg.fr (130.79.128.5) or via http://cdsweb.u-strasbg.fr/cgi-bin/qcat?J/A+A/.}}

   \author{Claudia Aguilera-G\'omez \inst{\ref{inst1}, \ref{inst2}} \and Iv\'an Ram\'irez \inst{\ref{inst3}} \and Julio Chanam\'e \inst{\ref{inst2},\ref{inst1}}
          }

   \institute{Millennium Institute of Astrophysics, Santiago, Chile \email{craguile@astro.puc.cl}\label{inst1} \and Instituto de Astrof\'isica, Pontificia Universidad Cat\'olica 
de Chile, Av. Vicu\~na Mackenna 4860, 782-0436 Macul, Santiago, Chile\label{inst2} \and Tacoma Community College, 6501 South 19th Street, Tacoma, WA 98466, USA\label{inst3}
 }

   \date{}

  \abstract
  % context heading (optional)
  % {} leave it empty if necessary  
   {}
  % aims heading (mandatory)
   {We address the existence and origin of the lithium (Li) desert, a region in the 
Li - $\mathrm{T_{eff}}$ plane sparsely populated by stars. Here we analyze some of the explanations that have been suggested for this region,  including mixing in the late main sequence, a Li dip origin for stars with low Li abundances in the region, and a possible relation with the presence of planets.}
  % methods heading (mandatory)
   {To 
study the Li desert, we measured the atmospheric parameters and Li abundance of 227 late-F dwarfs and subgiants, chosen to be in the $\mathrm{T_{eff}}$ range of the desert and without previous Li abundance measurements. Subsequently, we complemented those with 
literature data to obtain a homogeneous catalog of 2318 stars, for which we compute masses and ages. We characterize stars 
surrounding the region of the Li desert.}
  % results heading (mandatory)
   {We conclude that stars with low Li abundances below the desert are more massive and more evolved than stars above the desert. Given the unexpected presence of low Li abundance stars in this effective temperature range, we concentrate on finding their origin. We conclude that these stars with low Li abundance do not evolve from stars above the desert: at a given mass, stars with low Li (i.e., below the desert) are more metal-poor. }
  % conclusions heading (optional), leave it empty if necessary 
   {Instead, we suggest that stars below the Li desert are consistent with having evolved from the Li dip, discarding the need to invoke additional mixing to explain this feature. Thus, stars below the Li desert are not peculiar and are only distinguished from other subgiants evolved 
from the Li dip in that their combination of atmospheric parameters locates them 
in a range of effective temperatures where otherwise only high Li abundance 
stars would be found (i.e., stars above the desert).}

\titlerunning{Lithium abundance patterns of late-F stars}
   \maketitle
%
%-------------------------------------------------------------------

\section{Introduction}

Lithium (Li) is an element that can be used as a tool to study a wide variety 
of problems in astrophysics. For example, standard Big Bang nucleosynthesis 
predicts a primordial Li abundance of 
$\mathrm{A(Li)}$\footnote{$\mathrm{A(Li)}=log(N_{Li}/N_{H})+12.00$, where $N_x$ 
is the number of atoms of element ``$x$"}$=2.6$ \citep{Cyburt2008}, which 
disagrees strongly with the abundance found in metal-poor stars in the Galaxy 
\citep[\textit{Spite Plateau} of $\mathrm{A(Li)=2.2}$,][]{SpiteSpite1982}. 
Moreover, some kind of lithium enrichment is needed afterwards to reach the 
values found in meteorites, corresponding to $\mathrm{A(Li)}=3.26$ 
\citep{Lodders2009}. Thus, the Li abundance can also be used to study the 
chemical enrichment of the Milky Way, since the different mechanisms of Li 
production (Cosmic ray spallation, \citealt{ReevesFowlerHoyle1970}; core 
collapse supernovae, \citealt{Woosley1990}; novae, 
\citealt{ArnouldNorgaard1975, Tajitsu2015}; or in low mass stars, 
\citealt{CameronFowler1971} during the asymptotic giant branch 
\citealt{SackmannBoothroyd1992}, and possibly in the red giant branch as well 
\citealt{SackmannBoothroyd1999}) must be finely tuned to match the observed Li 
abundances of the Galaxy \citep[e.g.,][]{Prantzos2012, Prantzos2017}. And although the value 
found in meteorites is indicative of the Solar abundance at formation, it is 
significantly different from what is found in the Solar photosphere today, a 
value closer to $\mathrm{A(Li)}=1.05$ \citep{Asplund2009}; a difference of over 
two orders of magnitude that cannot be explained by standard stellar evolution. 

Li is burned by proton capture at temperatures and densities easily found at 
the interior of low-mass stars during the main sequence (MS) phase, but which, 
according to canonical models and helioseismic measurements \citep{CD1991}, are 
located below the convective envelope, thus predicting no Li changes during the 
MS. This puzzling observation indicates some mechanism of Li depletion is 
at work during the MS. Moreover, canonical stellar evolutionary models predict 
a certain Li abundance given the same mass, age, and metallicity of stars, which 
is in clear contrast with the wide spread in Li abundance found for stars of 
similar atmospheric parameters in the solar neighborhood, further stressing 
the need for a nonstandard mechanism of Li depletion acting inside low-mass 
stars. 

Among the candidate mechanisms for enhanced Li depletion during the MS we can find 
rotational mixing \citep{Pinsonneault1992, Eggenberger2010}, mass loss 
\citep{SwensonFaulkner1992}, diffusion \citep{Michaud1986}, gravity waves 
\citep{MontalbanSchatzman2000, CharbonnelTalon2005}, and overshooting 
\citep{XiongDeng2009}.

The problem of Li depletion in stars extends to different effective 
temperatures. Measurements in open clusters \citep[e.g., 
Hyades,][]{BoesgaardTripicco1986} show that F-dwarfs in a $300$ K temperature 
range around $6600$ K (corresponding to masses in the range $1.0 
-1.5\,\mathrm{M_\odot}$), are Li-depleted when compared to hotter and cooler 
stars by over $2$ dex. This pattern (The \textit{lithium dip}) has been 
confirmed to exist in field stars \citep{Randich1999,LambertReddy2004} as well 
as older and similarly aged open clusters (M34, \citealt{Jones1997}; NGC 752, 
\citealt{HobbsPilachowski1986}; M67 \citealt{Balachandran1995}, among others) 
but studies of F-dwarfs in younger open clusters like the Pleiades 
\citep{BoesgaardBudgeRamsay1988} were not able to find this signature, 
indicating that it must develop at a later age when the star is already in the MS phase \citep{SestitoRandich2005}.
Comparing the Li abundance of stars in different clusters, \citet{Balachandran1995} found that the location in mass of the Li dip depends on the cluster's metallicity, a correlation that was confirmed later on \citep[e.g.,][]{AnthonyTwarog2009}. Any attempt to find an explanation for the Li dip must not only take into 
account its morphology, but also the possible correlations between mass, age 
and metallicity of the stars \citep[e.g.,][]{DelgadoMena2015}. Such a consistent explanation has not been found so far. 

Another striking feature in the $\mathrm{A(Li)}$ - $\mathrm{T_{eff}}$ plane is 
the bimodal distribution shown by dwarfs of $\mathrm{T_{eff}}>5900$. First 
identified by \citet{Chen2001}, stars with high Li abundance are separated in 
this diagram from those with low Li abundances by a vacate area, the 
\textit{lithium desert}. The low abundance side of this desert was once thought 
to be connected to the Li dip, as stars with $\mathrm{T_{eff}}\sim 6600$ K 
could have evolved to populate that area. \citet{Chen2001} suggested this explanation based on the tight correlation between mass and metallicity that is found for the stars with low Li abundance in the region of the desert, that is also found for stars in the Li dip. However,  \citet{Ramirez2012} find that this tight correlation is also present for the high Li abundance stars. They propose that a process of Li depletion, taking place either during the MS or subgiant phases of stellar evolution, is responsible for the low Li abundance of stars below the Li desert. In this work, we find that both conclusions are partially correct, and that, although the stars with low Li abundances below the desert seem to be related to the Li dip, the correlation between mass and metallicity of stars is found regardless of Li abundance.

To this already complex picture of Li in low-mass stars, a new difficulty is to 
be added: the presence or absence of exoplanets. The possible interaction 
between the star and its protoplanetary disk during the pre-MS may slow down 
the surface rotation of the host star, increasing the differential rotation in 
its interior, enhancing the rotationally induced mixing that would reduce the 
Li abundance \citep{Bouvier2008}. Planet migration, that also affects the 
angular momentum of the host star could also have an effect on the Li abundance 
\citep{Castro2009}. This would indicate that MS stars without planets could 
present a higher Li abundance than their planet-hosting counterparts. 
Observationally, some works have claimed evidence of this enhanced depletion in 
planet hosts \citep{Israelian2004, Gonzalez2008, Israelian2009, 
DelgadoMena2014, Figueira2014, Gonzalez2015}, but this is still an ongoing 
debate, as several other studies have either found that there is no correlation 
or have attributed the alleged trend to biases in the observational samples 
\citep{Ryan2000, LuckHeiter2006, Baumann2010, Ghezzi2010,Ramirez2012, 
CarlosNissenMelendez2016}.
Confirming this trend could allow for selection of candidate host stars based on their 
Li abundances, while if no such trend exists, it would be indicative that the Li 
depletion is produced by some other noncanonical physics yet to be considered.

To best study the Li abundance and all the related problems introduced so far, 
a large, homogeneous sample of measurements is needed. So far, several works 
have presented internally consistent catalogs of Li abundance in dwarfs. The 
largest so far was recently presented by \citet{Guiglion2016} in the context of 
the AMBRE project, consisting of Li abundance for over $\sim7200$ stars. 
Although this catalog presents an exciting opportunity to study depletion and 
other Li-related phenomena, it does not allow us to reliably derive masses and 
ages for its stars (due to their lack of parallaxes), which are needed in most of these problems to derive a 
fully consistent picture. Samples with parallaxes to derive these parameters 
are ideal. The compiled catalog of \citet{Ramirez2012} is one of those, that 
used the derived atmospheric parameters for the observed stars and data form 
the literature to set all samples to the same scale, finally including $1381$ dwarfs and subgiants. Other works have presented 
new abundance measurements of FGK-dwarfs since then, that can also be 
homogenized to that scale, as long as they have parallaxes to calculate the ages 
and masses in a consistent way. Thus, the future of this kind of studies is in 
the \textit{Gaia} Mission, that will provide astrometric solution for a large number of stars in the Galaxy.

This paper extends the work of \citet{Ramirez2012} (R12 hereafter), following a 
similar procedure. We use this extended sample to study late F-stars and conclude about the existence, morphology, and origin of the Li desert. We describe the spectroscopic observations of stars in the 
temperature range of the Li desert and the determination of Li abundances and 
atmospheric parameters in Section \ref{SecObs}. We explain how the literature 
data was homogenized to create an extended catalog in Section \ref{catalog_sec} and Section 4 
presents a discussion on the nature of the lithium desert, focusing on the 
correlations of Li with mass, age, metallicity, evolutionary status, and 
finally relating its origin to the Li dip. We briefly discuss previous 
literature on the subject of the Li desert in Section \ref{disc}, and summarize 
and conclude in Section \ref{summ}.

\section{Observations and Li abundance analysis} \label{SecObs}
Since our primary goal is to investigate the lithium desert, we targeted dwarf and subgiant stars without previous Li abundance measurements potentially inside that $T_\mathrm{eff}$ range. We used a stellar parameter compilation catalog by J.\,Mel\'endez (private communication), excluded stars that were already in R12, and searched for objects with $V<8$ and $\mathrm{[Fe/H]}>-0.4$ in the $T_\mathrm{eff}$ range from 5800 to 6350 K. We allowed a wider $T_\mathrm{eff}$ range than that of the Li desert to account for systematic uncertainties in previously published stellar parameters. In some cases, we relaxed the $V$ magnitude restriction to allow somewhat fainter stars and prioritized the observation of targets that were known to host planets.

Spectroscopic observations were carried out in three runs at two locations: a five-night run at Las Campanas Observatory (LCO) with the MIKE spectrograph on the 6.5-m Clay-Magellan Telescope in July 2012, and two four-night runs at McDonald Observatory (McD) with the Tull spectrograph on the 2.7-m Harlan\,J.\,Smith Telescope in October 2012 and March 2013. We observed 89 stars from LCO and 139 stars from McD, in addition to obtaining a solar spectrum via reflected sunlight from asteroid Vesta at both locations. Details on the observational setup, data reduction, and post-reduction processing procedures are explained in detail in \citep[][, their Sect.~3.1]{Chaname2012} for the LCO data and in \citep[][, their Sect.~2.2]{Ramirez2014} for the McD spectra. Our spectra have a spectral resolution ($\lambda/\Delta\lambda$) of about 60\,000 and signal-to-noise ratios (S/Ns) well above 100 per pixel, with typical values around 400.

Stellar parameters and Li abundances for these 227 stars were measured as in R12. In short, temperatures were determined from colors and the color-temperature relations by \citet{Casagrande2010}. We used as many of the $B-V$, $b-y$, RI(C), Tycho, and 2MASS colors as available for each star. Revised {\it Hipparcos} parallaxes from \citet{vanLeeuwen2007} and theoretical isochrones were then employed to measure $\log\,g$ values (see Sect. \ref{catalog_sec} for more details) while [Fe/H] values were estimated from a line-by-line analysis using the solar spectrum (Vesta) as reference. The spectrum synthesis code MOOG \citep[][ 2014 version]{Sneden1973} was employed to calculate iron abundances for each spectral line in the star and Sun. The star minus Sun difference in abundance was first calculated for each line and the average of these differences is reported as our final iron abundance. This strictly differential method ensures higher precision in the elemental abundance measurement as line-dependent systematic uncertainties can be minimized in the process. The uncertainty in $T_\mathrm{eff}$ is dominated by the color-to-color scatter, which is quantified by the $1-\sigma$ standard deviation of the effective temperature values calculated for each color, while the uncertainty in $\log\,g$ depends primarily on the errors in the parallax. We propagated these errors into the elemental abundance measurements, assuming they are independent, and added the error contributions in quadrature along with the line-by-line scatter.

\begin{table}
\caption{Atmospheric parameters and Li abundances for the
 227 stars presented in this work \label{ourstars}}
 \centering
 \resizebox{0.49\textwidth}{!}{
 \begin{tabular}{lrrrr}
 \hline \hline
Name &$\mathrm{T_{eff}}$ (K)& $\log g$ & [Fe/H] & $\mathrm{A(Li)}$ \\
\hline
HIP682 & $5855\pm48$ & $4.39\pm0.03$ & $0.05\pm0.07$ & $3.09\pm0.05$ \\
HIP726 & $5986\pm37$ & $4.2\pm0.04$ & $0.24\pm0.05$ & $2.6\pm0.03$ \\
HIP937 & $6095\pm30$ & $3.85\pm0.06$ & $0.05\pm0.08$ & $1.92\pm0.02$ \\
HIP1444 & $5813\pm44$ & $4.26\pm0.03$ & $-0.09\pm0.05$ & $2.07\pm0.04$ \\
HIP1541 & $5781\pm30$ & $4.18\pm0.05$ & $0.04\pm0.05$ & $2.2\pm0.03$ \\
HIP1634 & $6311\pm42$ & $3.92\pm0.07$ & $0.11\pm0.06$ & $2.43\pm0.03$ \\
HIP2174 & $6046\pm24$ & $4.22\pm0.04$ & $0.18\pm0.04$ & $2.9\pm0.02$ \\
HIP2292 & $5956\pm33$ & $4.26\pm0.04$ & $0.19\pm0.04$ & $2.5\pm0.03$ \\
HIP4311 & $6235\pm61$ & $3.96\pm0.04$ & $-0.34\pm0.05$ & $1.45\pm0.04$ \\
HIP4393 & $5866\pm27$ & $4.4\pm0.02$ & $0.24\pm0.04$ & $0.85\pm0.02$ \\
\vdots & \vdots & \vdots & \vdots & \vdots\\
\hline
 \end{tabular}}
 \tablefoot{(Only a portion of Table \ref{ourstars} is shown for guidance 
on its format. The entire machine-readable table can be found electronically.)}
\end{table}

MOOG was also used to calculate the Li abundances of our sample stars using line profile fitting of the 6707\,\AA\ line region. Macroturbulent velocities were estimated from \citet{ValentiFischer2005} formula, while $V\sin\,i$ was solved as part of the spectrum fitting procedure. Errors in the lithium abundances are dominated by the uncertainties in stellar parameters, which were propagated in the process, assuming they are independent from each other, except in cases where the lithium line is very weak or not detected. For the latter, we employed a visual 2-$\sigma$ criterion, allowed by MOOG's residuals view, in which an upper limit was decided by ensuring that the majority of data points in the residuals stayed within 2-$\sigma$ of zero difference with the model spectrum. The spectra were normalized using IRAF's continuum task. The spectral order containing the lithium line in both the LCO and McD spectra is free from very strong lines, continuum determination using this standard tool is very accurate. Nevertheless, a pseudo-continuum was decided for each star during the line-fitting procedure to increase accuracy. In dwarf and sub-giant stars, the 6707\,\AA\ region of the spectrum contains several small windows free from strong lines that one can rely on for this procedure. In both [Fe/H] and Li abundance measurements, we employed MARCS model atmospheres with standard chemical composition \citep{Gustafsson2008}, interpolated linearly to the stellar parameters of each problem star.

Given that the quality of our new spectroscopic data and the methods of stellar parameter and lithium abundance determination are identical to those in R12, on average the uncertainties that we derive here are also very similar: $\Delta T_\mathrm{eff}\simeq50$\,K, $\Delta\log\,g\simeq0.06$\,dex, $\Delta\mathrm{[Fe/H]}\simeq0.05$\,dex, and $\Delta A\mathrm{(Li)}\simeq0.05$\,dex. The stellar parameters and Li abundances for the 227 stars we observed specifically for this study are given in Table \ref{ourstars}.

\section{Assembling the catalog} \label{catalog_sec}
The final catalog of Li abundances includes our new measurements for stars in 
the temperature range of the Li desert (Section \ref{SecObs}), and literature 
data from \citet{DelgadoMena2014,DelgadoMena2015} and 
\citet{Gonzalez2014,Gonzalez2015}. Several other works have reported Li 
abundances since 2012, but we concentrate on large available samples of field 
stars. The catalog also includes the compilation already done in R12 which 
includes previous measurements of Li abundances in dwarfs and subgiants from 
\citet{LambertReddy2004}, \citet{LuckHeiter2006}, \citet{Israelian2009}, 
\citet{Baumann2010}, \citet{Ghezzi2010}, \citet{Gonzalez2010} and 
\citet{Takeda2010}. In turn, \citet{LambertReddy2004} contains parameters and 
abundance measurements from \citet{Balachandran1990}, \citet{Chen2001} and 
\citet{Reddy2003}. All of these works have high-quality data and provide 
internally consistent measurements.

To consider systematic differences in the calculation of stellar parameters or 
Li abundances, we normalize the literature data to ours by following a similar 
procedure to R12. Atmospheric parameters and Li abundances obtained in Section 
\ref{SecObs} are internally consistent with those found in R12, so for the 
purpose of homogenizing the catalog, we consider the 898 stars (671 from R12 
and 227 presented in this work) to be part of a larger sample used to 
normalize the remaining literature data.

\begin{table*}
\caption{Calculated offsets in atmospheric parameters and Li abundances 
between literature data and this work plus R12. \label{offsets}} 
\centering
\begin{tabular}{lrrrrrrr}
\hline \hline
Sample\tablefootmark{a} & $\Delta \mathrm{T_{eff}}$ (K) & $\Delta \log g$ & $\Delta$[Fe/H] & $\Delta \mathrm{A(Li)}$ & $\mathrm{N_{c}}$\tablefootmark{b}& $\mathrm{N_{u}}$\tablefootmark{c} & $\mathrm{N_{o}}$\tablefootmark{d}\\
\hline
TW+R12 & $  0\pm 00$ & $ 0.00\pm0.00$ & $ 0.00\pm0.00$ & $ 0.00\pm0.00$ & - & 898 & 835 \\
B10 & $ 40\pm48$ & $ 0.05\pm0.06$ & $ 0.03\pm0.03$ & $ 0.07\pm0.09$ &  69 & 117 & 117 \\ 
G10 & $ -3\pm69$ & $-0.02\pm0.04$ & $ 0.04\pm0.04$ & $-0.14\pm0.12$ &  45 & 138 & 152 \\
I09 & $ 26\pm32$ & $ 0.05\pm0.05$ & $ 0.02\pm0.02$ & $ 0.06\pm0.04$ &  11 &  79 & 80 \\
LH06 & $ 46\pm91$ & $ 0.04\pm0.20$ & $ 0.01\pm0.06$ & $ 0.01\pm0.11$ &  77 & 165 & 194 \\
T10 & $-12\pm54$ & $-0.03\pm0.09$ & $ 0.03\pm0.04$ & $-0.03\pm0.10$ &  26 & 113 & 117 \\
Gh10 & $ 30\pm71$ & $-0.06\pm0.14$ & $-0.00\pm0.06$ & $ 0.01\pm0.08$ &  38 & 253 & 262 \\
LR04 & $-83\pm63$ & $-0.01\pm0.11$ & $-0.02\pm0.05$ & $-0.07\pm0.08$ & 212 & 415 & 451 \\
DM14 & $  7\pm80$ & $ 0.07\pm0.09$ & $ 0.01\pm0.04$ & $ 0.04\pm0.12$ &  35 & 307 & 326 \\
DM15 & $ 46\pm70$ & $ 0.12\pm0.12$ & $ 0.03\pm0.06$ & $ 0.04\pm0.10$ &  65 & 778 & 836 \\
G1415 & $ 42\pm82$ & $ 0.01\pm0.04$ & $ 0.02\pm0.05$ & $-0.17\pm0.07$ &  29 &  66 & 68 \\
\hline
\end{tabular}
\tablefoot{
\tablefoottext{a}{Samples correspond to: B10=\citet{Baumann2010}; 
G10=\citet{Gonzalez2010}; I09=\citet{Israelian2009}; 
LH06=\citet{LuckHeiter2006}; T10=\citet{Takeda2010}; Gh10=\citet{Ghezzi2010}; 
LR04=\citet{LambertReddy2004}; DM14=\citet{DelgadoMena2014}; 
DM15=\citet{DelgadoMena2015}; G1415=\citet{Gonzalez2014}+\citet{Gonzalez2015};  TW=This work}
\tablefoottext{b}{Number of stars in common between the sample and data from 
this work plus R12.}
\tablefoottext{c}{Number of stars from this sample used in the final catalog (after removal of possible spectroscopic binaries and stars without parallaxes).}
\tablefoottext{d}{Total number of stars in the original catalog, removing giants when necessary.}
}
\end{table*}

\begin{table*}
\caption{Catalog of homogenized stellar parameters, lithium abundances and 
calculated masses and ages. A flag for planet hosting status is also included. 
\label{catalog}} 
\centering
\begin{tabular}{crrrrrrrc}
\hline \hline
Name & $\mathrm{T_{eff}}$ (K) & $\log g$ & [Fe/H] & $\mathrm{A(Li)}$ & Mass ($\mathrm{M_\odot}$) & Age (Gyr) \tablefootmark{a}& Planet & 
Source\tablefootmark{b}\\
\hline
HIP57 & $5138\pm100$ & $4.52\pm0.03$ & $-0.06\pm0.08$ & $<0.02$ & $0.80^{+0.05}_{-0.01}$ & $13.14^{+0.31}_{-8.13}\mathrm{\ [iso]}$ & no & DM15 \\
HIP80 & $5886\pm80$ & $4.13\pm0.08$ & $-0.54\pm0.08$ & $1.76\pm0.12$ & $0.92^{+0.04}_{-0.04}$ & $10.11^{+1.72}_{-1.14}\mathrm{\ [iso]}$ & no & DM14 \\
HIP348 & $5740\pm50$ & $4.39\pm0.05$ & $-0.17\pm0.04$ & $0.85\pm0.07$ & $0.94^{+0.03}_{-0.02}$ & $7.66^{+2.15}_{-2.68}\mathrm{\ [iso]}$ & (...) & R12+B10 \\
HIP394 & $5636\pm50$ & $3.76\pm0.07$ & $-0.49\pm0.04$ & $1.98\pm0.02$ & $1.10^{+0.15}_{-0.05}$ & $5.00^{+2.28}_{-1.05}\mathrm{\ [iso]}$ & (...) & R12+LR04 \\
HIP413 & $6134\pm100$ & $4.35\pm0.05$ & $-0.12\pm0.10$ & $2.61\pm0.10$ & $1.11^{+0.03}_{-0.06}$ & $3.51^{+2.12}_{-1.49}\mathrm{\ [iso]}$ & no & DM15 \\
HIP459 & $5738\pm100$ & $4.52\pm0.04$ & $0.02\pm0.10$ & $1.87\pm0.12$ & $0.96^{+0.03}_{-0.04}$ & $1.96^{+4.64}_{-0.66}\mathrm{\ [iso]}$ &  no & DM14 \\
HIP475 & $5836\pm72$ & $4.36\pm0.05$ & $-0.34\pm0.06$ & $1.63\pm0.05$ & $0.92^{+0.03}_{-0.03}$ & $8.75^{+2.73}_{-1.64}\mathrm{\ [iso]}$ & (...) & R12 \\
HIP493 & $5943\pm50$ & $4.38\pm0.03$ & $-0.23\pm0.04$ & $2.31\pm0.02$ & $0.99^{+0.03}_{-0.02}$ & $5.77^{+1.13}_{-2.00}\mathrm{\ [iso]}$ &(...) & R12+LR04 \\
HIP522 & $6277\pm50$ & $4.24\pm0.02$ & $0.04\pm0.04$ & $2.85\pm0.02$ & $1.24^{+0.01}_{-0.01}$ & $2.93^{+0.36}_{-0.37}\mathrm{\ [iso]}$ & yes & R12+Gh10+DM15 \\
HIP530 & $5868\pm50$ & $3.91\pm0.05$ & $-0.05\pm0.04$ & $0.70\pm0.03$ & $1.22^{+0.06}_{-0.04}$ & $5.15^{+0.54}_{-0.70}\mathrm{\ [iso]}$ &(...) & R12+LR04 \\
\vdots & \vdots & \vdots & \vdots & \vdots & \vdots & \vdots & \vdots & \vdots\\
\hline
\end{tabular}
\tablefoot{(Only a portion of Table \ref{catalog} is shown for guidance 
on its format. The entire machine-readable table can be found electronically.)}
\tablefoottext{a}{This columns also includes the age determination method. ``iso" indicates ages derived through isochrone fitting and ``rot" ages obtained with gyrochronology.}
\tablefoottext{b}{Sources correspond to: B10=\citet{Baumann2010}; 
G10=\citet{Gonzalez2010}; I09=\citet{Israelian2009}; 
LH06=\citet{LuckHeiter2006}; T10=\citet{Takeda2010}; Gh10=\citet{Ghezzi2010}; 
LR04=\citet{LambertReddy2004}; DM14=\citet{DelgadoMena2014}; 
DM15=\citet{DelgadoMena2015}; G1415=\citet{Gonzalez2014}+\citet{Gonzalez2015}; 
R12=\citet{Ramirez2012}; This=This work.}
\end{table*}

The homogenization process uses the stars in common between our data (including 
the sample from R12) and previously available samples to compute offsets in 
$\mathrm{T_{eff}}$, $\log g$, [Fe/H], and lithium abundance. Stars with 
lithium upper limits are excluded from the calculation of offsets, but their 
abundances are still homogenized following the same procedure used for 
detections. Then, these offsets are subtracted to literature data to homogenize 
the catalog. Calculated offsets can be found in Table \ref{offsets}.

We repeated the homogenization process using different subsamples to see how the resulting offsets and parameters change. Although the focus of this work is late-F dwarfs, using these stars as a comparison sample is not the best option, as some of the previously available samples only have solar-type stars and thus we find no stars in common with our work. Nevertheless, for those catalogues where homogenization is possible, we find very similar offsets to those presented in Table \ref{offsets} within the given uncertainties. The same can be said when we use solar-type stars as a comparison sample, and, although parameters can be different when using a different comparison sample, they are always well within the error bars. To have a larger number of stars in common between previously available catalogs and our measurements, we have decided to use the entire sample in the homogenization process.

A detailed description of the process of homogenization and a discussion on the 
first compilation can be found in R12. Here we summarize some important points:
\begin{itemize}
\item When non-LTE corrections were applied to the Li abundance measurements, 
they were reverted, and we recovered the LTE abundances.
\item In case a star was present in multiple samples, we take a weighted 
average for the atmospheric parameters and Li abundances, where the weights are 
the variances in each work.
\item Errors in \citet{Israelian2009} and \citet{DelgadoMena2014, 
DelgadoMena2015} are relatively small when compared to other samples. Even if these errors are internally consistent, systematic error dominates when comparing with other works. To give a 
similar weight to their data and data from other sources, their errors were increased (see R12 for a detailed description of this process).
\item The change in stellar parameters when we do not weight the average-taking 
procedure is negligible.
\item When an upper limit is given for the Li abundance in more than one 
source, the lowest upper limit is considered.
\item When there is an upper limit and a detection for the Li abundance of a star, the detection is considered.
\end{itemize}

We obtain a preliminary catalog of 2484 stars with homogeneous 
$\mathrm{T_{eff}}$, $\log g$, [Fe/H] and $\mathrm{A(Li)}$. 

We remove known spectroscopic binaries from our catalog, given that stars in close binary systems can interact and thus their Li evolution may be different from that of single stars. Also, the photometry, which we need in our catalog to obtain the absolute magnitude of the stars, could be compromised for these binary systems. We cross-match our catalog of 2484 stars with the Ninth Catalogue of Spectroscopic Binary Orbits \citep{Pourbaix2004} and find 96 spectroscopic binaries. We stress that these are only the spectroscopic binaries and it is still possible to have other types of binary contamination in the sample.

Unfortunately, not all of the resulting 2388 stars can be used in our final compilation, since we will need to derive masses and ages for the stars by using theoretical isochrones. To do this we use absolute magnitudes which require parallaxes for the stars. Stars without parallaxes are removed from the final catalog. In this work, \textit{Hipparcos} parallaxes are used \citep{vanLeeuwen2007}. The joint \textit{Tycho-Gaia} Astrometric Solution (TGAS), as a part of the \textit{Gaia} Data Release 1 \citep{Gaiacoll2016}, became available during the development of this work, but about a quarter of the stars in our sample are not part of TGAS. For the 1857 stars that had data from both sources, most of the parallaxes are consistent. The exceptions are 18 stars, for which \textit{Hipparcos} and \textit{Gaia}/TGAS parallaxes are very different. We have found that most of these stars are peculiar and probable binaries, and as we cannot be sure about the determination of their parallaxes, we remove them from the final sample. 

After discarding the spectroscopic binaries, stars with nonmatching \textit{Hipparcos} and \textit{Gaia DR1} parallaxes, and stars without parallaxes (most of which are stars with planets presented in \citealt{DelgadoMena2014,DelgadoMena2015} without HARPS measurements), 2318 stars remain, which are used for further analysis. 

To calculate masses, ages, and $\log g$, stars with parallaxes are placed on a color-magnitude diagram (CMD) by using their $\mathrm{T_{eff}}$ and absolute magnitudes, and their positions are compared with 
theoretical Yonsei-Yale isochrones \citep{Yi2001, Demarque2004}. Several different combinations of stellar parameters ($log g$, [Fe/H], mass, and age) can produce isochrones that go through the same point in the CMD. Thus, we use metallicities as obtained with spectra to find the age of the star.
We construct probability distributions for the stellar age, mass and $\log g$ with all the 
isochrone points in a $3\sigma$ radius from the observed parameters, and take into account the uncertainties on these parameters. The method that uses the probability distribution to calculate the stellar age also considers how isochrones cluster in the CMD, and calculates errors accordingly. A much more detailed discussion on the errors associated with this process to calculate parameters based on isochrones can be found in \citet{Nordstrom2004}. Subsequently, the most 
probable parameters are adopted, along with the $1\sigma$ Gaussian-like errors. 
More details about this widely used process can be found in 
\citet{Chaname2012}. It is worth noting that as the age distributions are 
asymmetric, errors in age are also asymmetric.

The $\log g$ obtained from the isochrone fitting process is preferred over the 
value derived from spectroscopy (given by each work separately) given that they are directly derived from the $\mathrm{T_{eff}}$ using the same method for each star and, as such, are more internally consistent. For most of the dwarfs and 
subgiants analyzed, the ages obtained through this process are fairly well 
constrained, but this is not the case for younger stars, for which rotation 
periods can be obtained to estimate the stellar age using gyrochronology relations \citep[][]{Barnes2003, Barnes2007, MamajekHillenbrand2008}. This method uses the idea that the rotation of stars with a convective envelope decline with stellar age, and as such, using 
the measured rotation period and a relation between period and age 
\citep{MamajekHillenbrand2008}, we can estimate the ages. In this work, as was done 
previously by R12, we use rotation periods from \citet{Gaidos2000}, 
\citet{Strassmeier2000}, and \citet{Pizzolato2003}. We consider ages obtained through rotation only up to $2.5$ Gyr, given the lack of a reliable calibration of gyrochronology relations for older stars. A total of 64 stars have ages obtained through this method.

The two different methods used to calculate ages give different typical errors. For gyrochronology, uncertainties are typically of $0.1\,\mathrm{Gyr}$ or $20\%$ of the age value, while the isochrone fitting method gives very different errors. These are calculated from the age probability distribution function, where the peak is the reported age. This function is not symmetrical, so different $1-\sigma$ lower and upper errors are adopted. Typically, errors tend to range between $1.0$ and $2.5\,\mathrm{Gyr}$, but they can be as large as $9.0\,\mathrm{Gyr}$.
We define stars with reliable ages as those with age/(error age)$>3.0$. When errors are asymmetrical, this condition has to be fulfilled for both lower and upper errors. The total number of stars with reliable ages are 1093, or $\sim 47\%$ of the sample.

 Although we have already discussed how \textit{Gaia}/TGAS and \textit{Hipparcos} parallaxes are very similar for most of the stars, we also analyze how the smaller parallax errors of TGAS impact the determination of masses, ages, and their errors. Masses calculated using both sets of parallaxes tend to be similar, even more so when the calculated masses are small. The errors in mass with both sets of parallaxes, are generally small, below $0.1\,\mathrm{M_\odot}$. In contrast, age can be very uncertain, even when using the more precise TGAS parallaxes. For both mass and ages, errors are usually similar or smaller when using TGAS parallaxes, but this is not always the case. The analysis and conclusions presented in the following sections, which use \textit{Hipparcos} parallaxes, do not change when using data from \textit{Gaia}/TGAS instead.

\begin{figure}[!hbt]
\begin{center}
\includegraphics[width=0.4\textwidth]{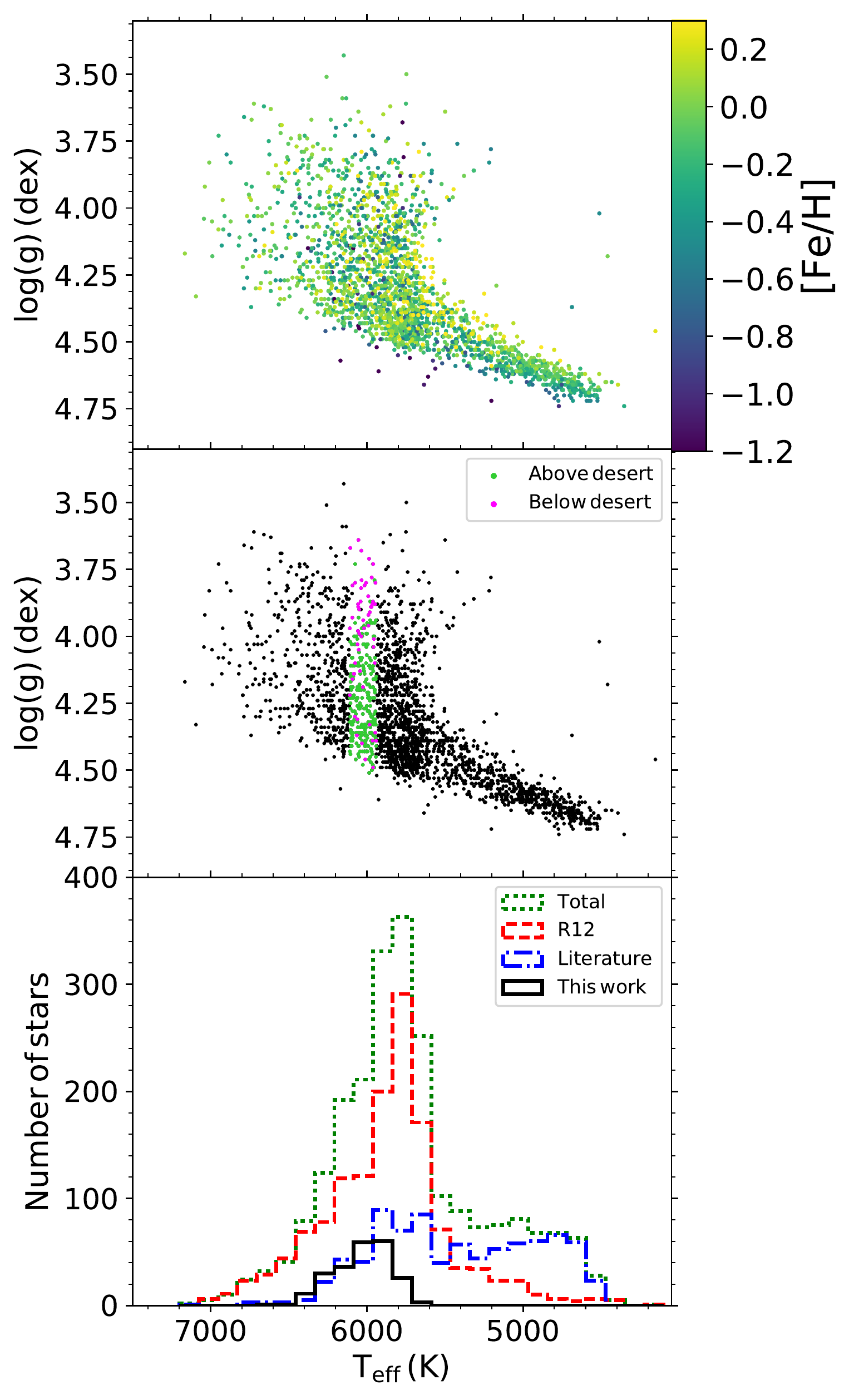}
\caption{Top panel  and middle panel: Stars of the sample in the $\mathrm{T_{eff}}$-$\mathrm{\log g}$ plane. Stars in the top panel are color-coded by their metallicities, and in the middle panel, stars in the effective temperature region of the Li desert are color-coded according to their Li abundance. Bottom panel: Temperature distribution of the final catalog compared to the 
compilation of \citet{Ramirez2012}, the largest sample with measured stellar 
ages and masses so far (R12), the new literature data obtained from 
\citet{DelgadoMena2014, Gonzalez2014, DelgadoMena2015} and \citet{Gonzalez2015} 
(Literature) and the new measurements provided by this work (This work). We note that, while few in relative terms, the new measurements reported in this work span, by design, the temperature region of the Li desert. }
\label{histtemp}
\end{center}
\end{figure}

We show our entire sample in Figure \ref{histtemp}. The top panel shows a typical $\mathrm{T_{eff}}$-$\mathrm{\log g}$ diagram, where we can identify stars in different evolutionary phases: MS, turn-off, and subgiant. These are color-coded by their metallicity. In the middle panel of the same figure, only stars in the effective temperature range of the Li desert are color-coded by their Li abundances. We discuss this feature in detail, and the differences between stars above and below the Li desert, in Section \ref{SecLiDesert}. The bottom panel of Figure \ref{histtemp} shows the temperature distribution of our final catalog, 
compared to the R12 compilation, the new literature data included since then 
\citep{DelgadoMena2014, DelgadoMena2015,Gonzalez2014,Gonzalez2015}, and the 
observations presented in this work. In particular, \citet{DelgadoMena2015} 
present a sample of very cool stars with mostly Li abundance upper limits which 
can be seen as a peak at lower temperatures in the temperature distribution of 
new literature data, and that makes most of the final catalog in this region. 
In comparison, for solar-like stars there is a significant contribution from 
every source, and for temperatures around the lithium desert 
($\mathrm{T_{eff}}\sim5950-6100$ K), there is a large increase in statistics 
given by our new observations.

We also attempted to apply corrections for non-LTE (NLTE) to the Li abundances of our sample using interpolation routines by \citet{Lind2009}. Knowledge of the microturbulence velocity is required to obtain these corrections, so we estimate it based on stellar effective temperature and $\log g$, following the relation by \citet{Bruntt2012} and the relation by M. Bergemann and V. Hill used for the Gaia-ESO Survey \citep{Gilmore2012}. 
We are unable to calculate NLTE corrections for 821 stars that fall outside the range covered by the grid. The NLTE corrections for 1285 stars are smaller than $\mathrm{\Delta A(Li)_{NLTE}=0.05}$, within the errors in measurements. We have repeated the analysis presented in this work using the corrected Li abundances and our conclusions do not change. Moreover, most of the analysis requires comparisons between stars with similar atmospheric parameters, which have similar NLTE corrections. Given that the sample with NLTE corrections applied is much smaller, but the correlations and results found are identical, we decide to use only the LTE Li abundances in this work, and we do not report the NLTE abundances.

To complete our catalog, we also include a flag for each star to indicate its 
substellar-mass-companion-hosting status: known planet host, no planets detected so far even after 
radial velocity follow up, or unknown. The compilation of planet hosting 
conditions is based on data from The Exoplanet Orbit Database \citep{Han2014}, 
\citet{FischerValenti2005}, \citet{Sousa2008}, \citet{Ghezzi2010}, 
\citet{Brugamyer2011}, and \citet{DelgadoMena2014, DelgadoMena2015}. 
According to these works, 225 stars are planet hosts, while 1245 are, to this 
date, thought to have no planets orbiting them. In cases where a star was present 
in more than one table, we include the result of the last work, which generally 
means a planet has been found around the star.

Our final compiled catalog can be found in Table \ref{catalog}, where we show 
the stellar parameters, derived ages and masses, and the planet flag based on 
the procedure described immediately above.

Trends of Li abundance with other stellar parameters in our sample and an overview of the catalog can be found in Appendix \ref{overview}, where we also refer to the age trend for solar twins in the sample and analyze the planet-hosting status of the stars.

\section{The lithium desert} \label{SecLiDesert}

\begin{figure}
\begin{center}
\includegraphics[width=0.5\textwidth]{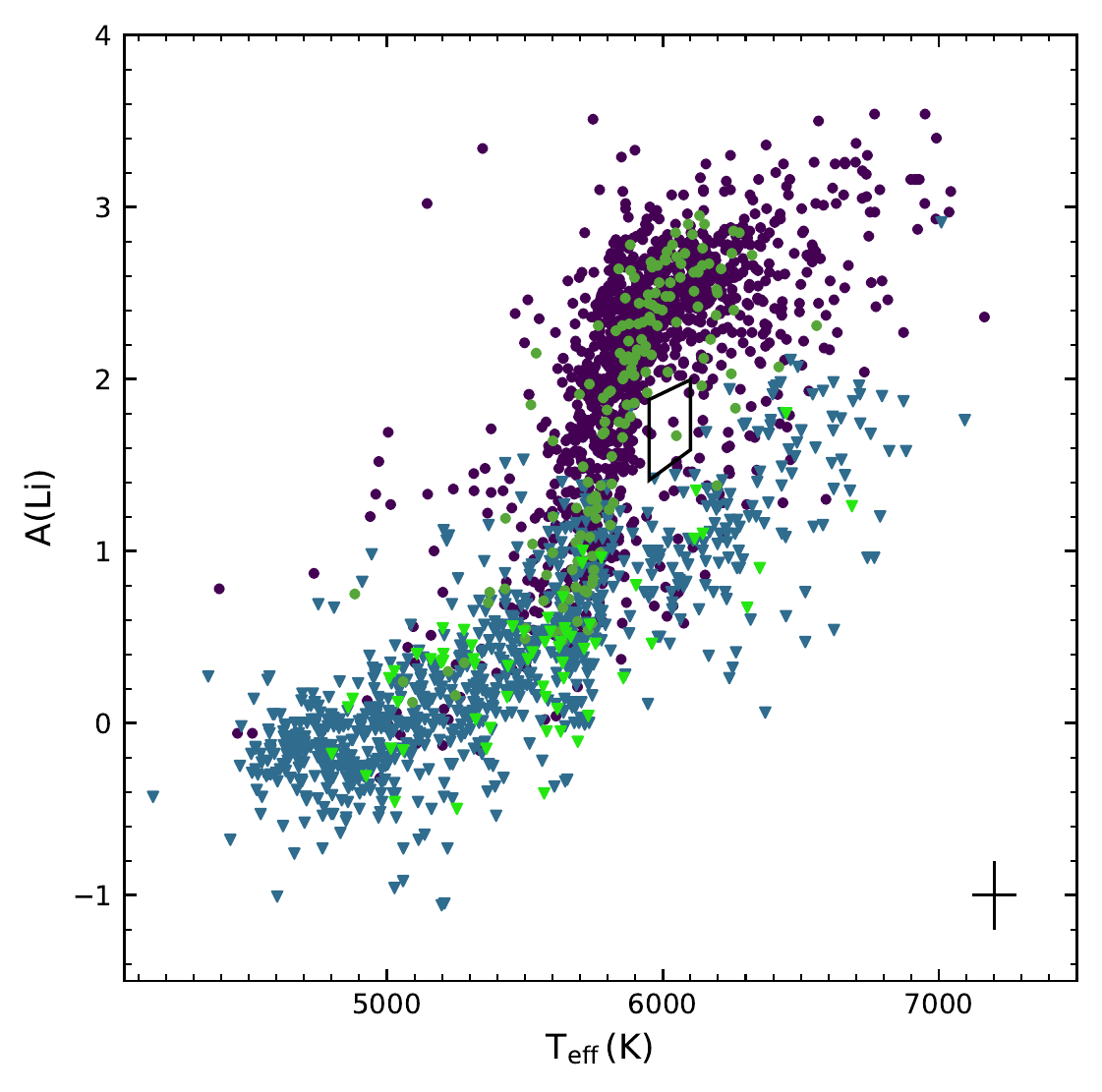}
\caption{Li abundance of stars in the final catalog, as a function of effective 
temperature. Circles (triangles) represent Li abundance determinations (upper limits), and the 
sources are also color-coded according to their planet hosting status. Green 
points are planet hosts, while blue points are the comparison sample, including 
non-hosts and stars with unknown status. A polygon is drawn to 
show the approximate location of the lithium desert.}
\label{AllvsTeff}
\end{center}
\end{figure}

Figure \ref{AllvsTeff} presents the Li abundance of stars in the compiled 
catalog and their distribution with $\mathrm{T_{eff}}$. Even with the addition of the new stars provided in this work, which exclusively span the region of the Li desert, this feature is apparent in Figure \ref{AllvsTeff} between $5950$ and $6100$ K. An arbitrary polygon (the same as seen in R12) is drawn here only to guide the eye, and does not exactly indicate the position of the desert. 

Other expected features can also be identified in this plot: There is a positive correlation between the Li upper limits with $\mathrm{T_{eff}}$, that is artificially produced by the detection limits of the Li doublet, i.e., at constant S/N, the Li lines are stronger for lower temperatures. Also, near the $\mathrm{T_{eff}}$ of the Sun ($5500\lesssim \mathrm{T_{eff}}\lesssim 6000$ K), cooler stars that have larger convective envelopes 
have lower lithium abundances than their hotter counterparts. However, at each temperature, there are stars that have low lithium abundances when compared to expectations from canonical stellar evolution \citep[e.g.,][]{Pinsonneault1997}, confirming that a Li depletion mechanism is 
acting on some of them while on the MS. Further correlations of Li with other parameters and an overview of the catalog can be found in Appendix \ref{overview}. Generally, the patterns found in R12 are confirmed using our larger sample.

\begin{figure*}[!hbt]
\centering
\includegraphics[trim={0 1.5cm 0.5cm 0}, clip=true, width=0.24\linewidth]{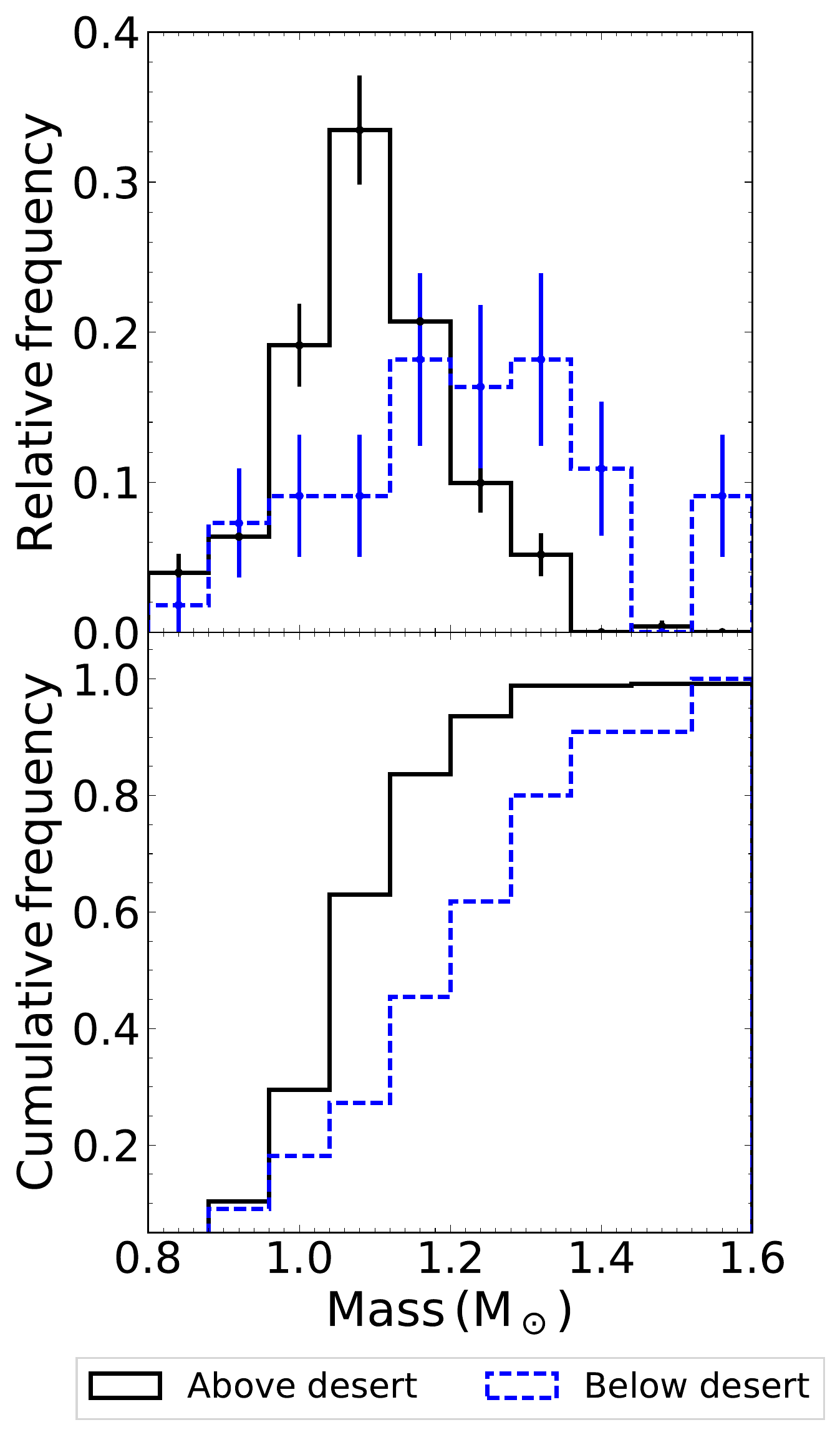}\includegraphics[trim={0 1.5cm 0.5cm 0}, clip=true, width=0.24\linewidth]{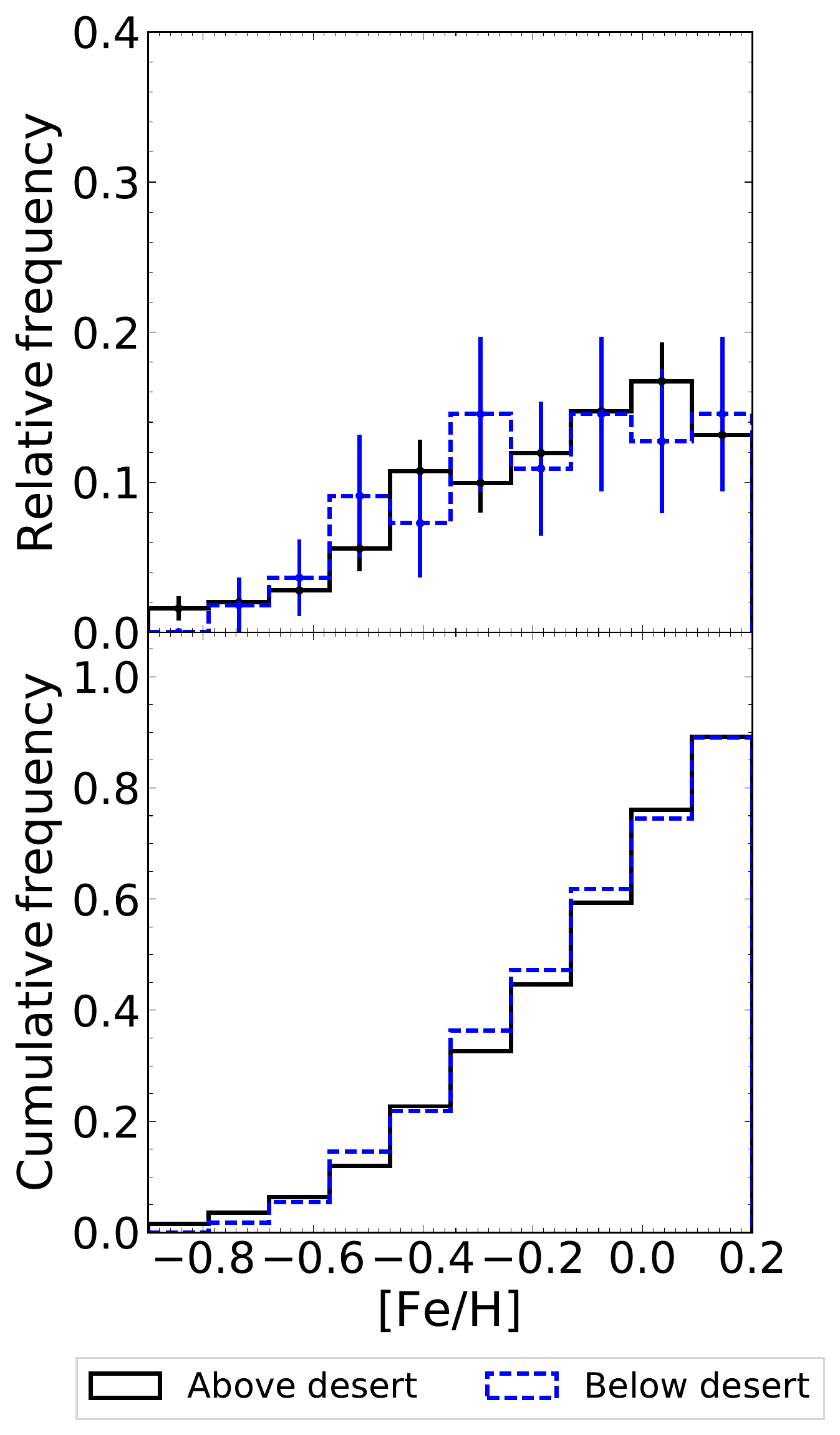}\includegraphics[trim={0 1.5cm 0.5cm 0}, clip=true, width=0.24\linewidth]{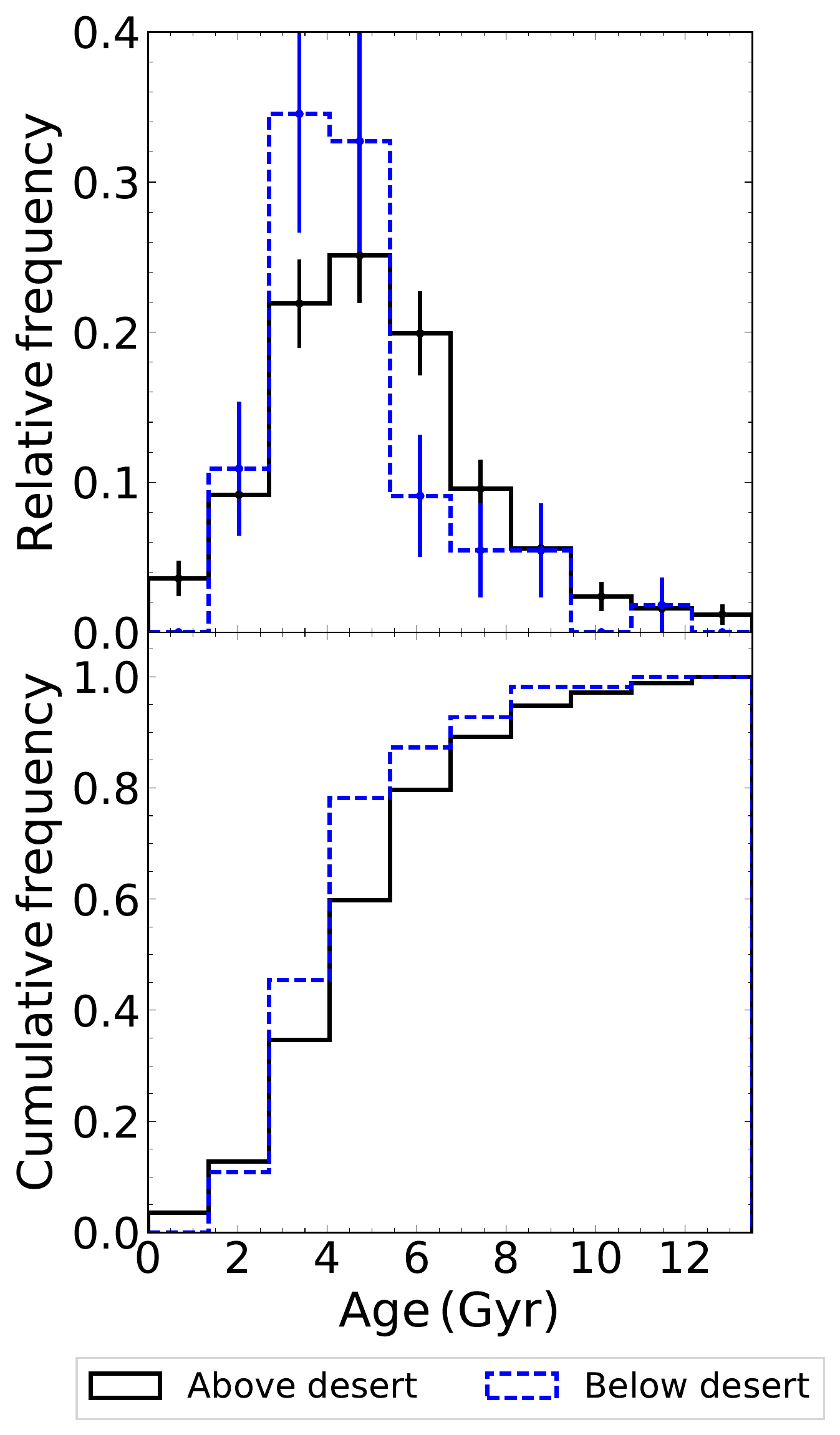}\includegraphics[trim={0 1.5cm 0.5cm 0}, clip=true, width=0.24\linewidth]{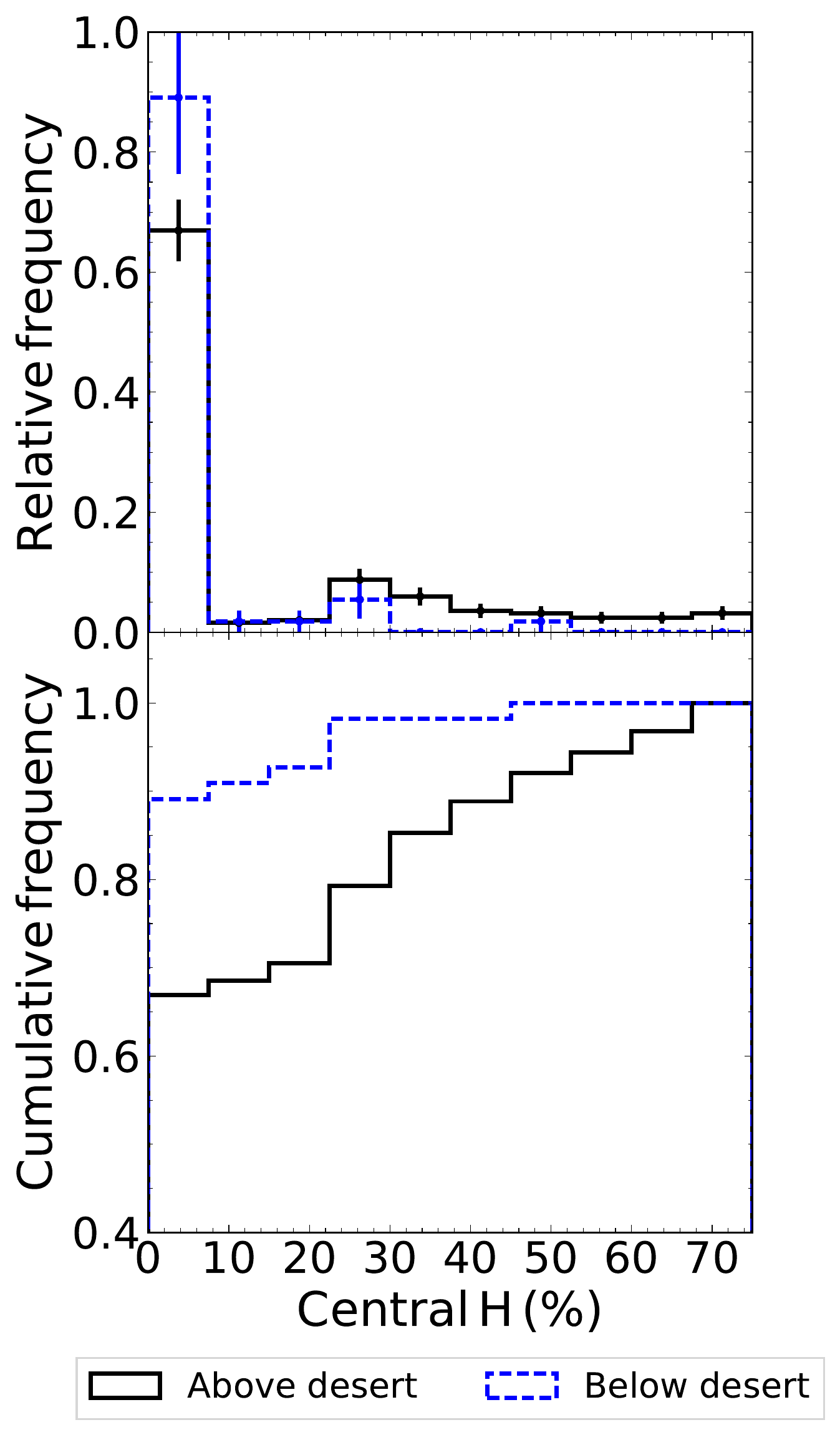}\\
\includegraphics[trim={0 0 0 22cm}, clip=true, width=0.5\linewidth]{Mass_histo_cum.pdf}
\caption{From left to right, distributions of mass, metallicity, age, and hydrogen percentage in the core for stars below (blue dashed line) and above (solid line) the desert. Top and bottom panels show the frequency and cumulative distributions, respectively. The distributions of metallicity and age look very similar for both groups of stars, in contrast to the mass and core hydrogen that seem to be different for stars with high or low Li content. \label{Hcentral}}
\end{figure*}

Studies so far about the Li desert and possible correlations of stars 
surrounding it have provided certain clues about the nature of this region, but 
this feature is still not completely understood. 
R12 noticed that there are stars with both high and low Li abundances at specific mass ranges (see Fig. 2 in R12 or the $1.1\,\mathrm{M_{\odot}}$ panel of Fig. \ref{Ap1} in Appendix \ref{charac} of this paper for an example). Stars with the same stellar parameters are expected to follow a similar evolution, and thus have similar Li abundances. However, as R12 find stars of different Li content at the same mass, they suggest that there might be some mechanism depleting the superficial Li abundance of stars during the MS or subgiant phases. Moreover, the small number of stars in the region of the desert may suggest that if a depletion mechanism is in action, it may be very short-lived and dramatic, producing a fast movement of stars in the $\mathrm{A(Li)}-\mathrm{T_{eff}}$ plane from high to low Li abundances.
This depletion mechanism is not the only possible explanation for the stars below the Li desert. It has also been suggested that low Li abundance stars in this region are related to the Li dip \citep{Chen2001}.

Based on these suggestions, we carefully analyze the Li desert using our compiled 
catalog. Although this feature remains in the 
$\mathrm{A(Li)}$-$\mathrm{T_{eff}}$ plane, it loses some clarity when compared 
to previous works with smaller samples because of the presence of a few stars 
in the previously empty region. It is possible that the desert extends to higher temperatures as well: there is also a small amount of stars from $\mathrm{T_{eff}}\sim6100$ K up to $\sim6300$ K. It is important to note that the lower side of the Li desert is mainly composed of upper 
limits, which means that the desert may extend to even lower Li abundances. Therefore, the Li desert may actually only separate between stars that do have Li in their atmosphere and those that do not.

Figure \ref{AllvsTeff} is also color-coded according to the presence of planets. Green points are planet hosts, and blue points are the comparison sample. One of the suggestions of R12 is that in the effective temperature range of the desert, planet hosts are only located in the upper side, with higher Li abundances (See, panel Da of their Fig. 9). We find one planet-hosting star in the lower side of the desert and one planet-hosting star located in the position of the desert. Thus, although we cannot completely discard the hypothesis that the presence of planets may be linked to the final Li abundance of the stars, the lack of planet-hosting stars with low Li is probably more related to the lower amount of stars below the Li desert (251 stars above the desert compared to 55 below), or the different distribution of stellar parameters above and below the desert and how this relates to the presence of planets, and not to a physical process.

Two stars are now identified to be in the region originally defined as the lithium desert. 
These are HIP22826 and HIP51028, the former being a planet-host. Besides their 
position in the $\mathrm{T_{eff}}$-$\mathrm{A(Li)}$ plane, these stars do not 
have any special or distinguishing characteristic. 
These are not the first stars found in that region. \citet{LopezValdivia2015} 
have identified one star (BD+47 3218) that falls in the desert. The data of \citet{LopezValdivia2015} are consistent with the previously 
compiled catalog of R12, so we can assume that BD+47 3218 is fairly well 
positioned in the diagram. Despite the presence of these few stars in this region, 
the Li desert does not seem to be simply a product of selection effects of the sample or 
uncertainties in measurements, especially when considering the sample size.

We address two specific questions regarding the Li desert in this section. First, we fully characterize the region, mapping the different distribution of stellar parameters above and below the desert. Then, we address the origin of this desert based on the previous characterization, considering how all these parameters can affect the Li abundance in the $\mathrm{T_{eff}}$ range studied.

\begin{figure*}[!hbt]
\begin{center}
\includegraphics[width=0.7\textwidth]{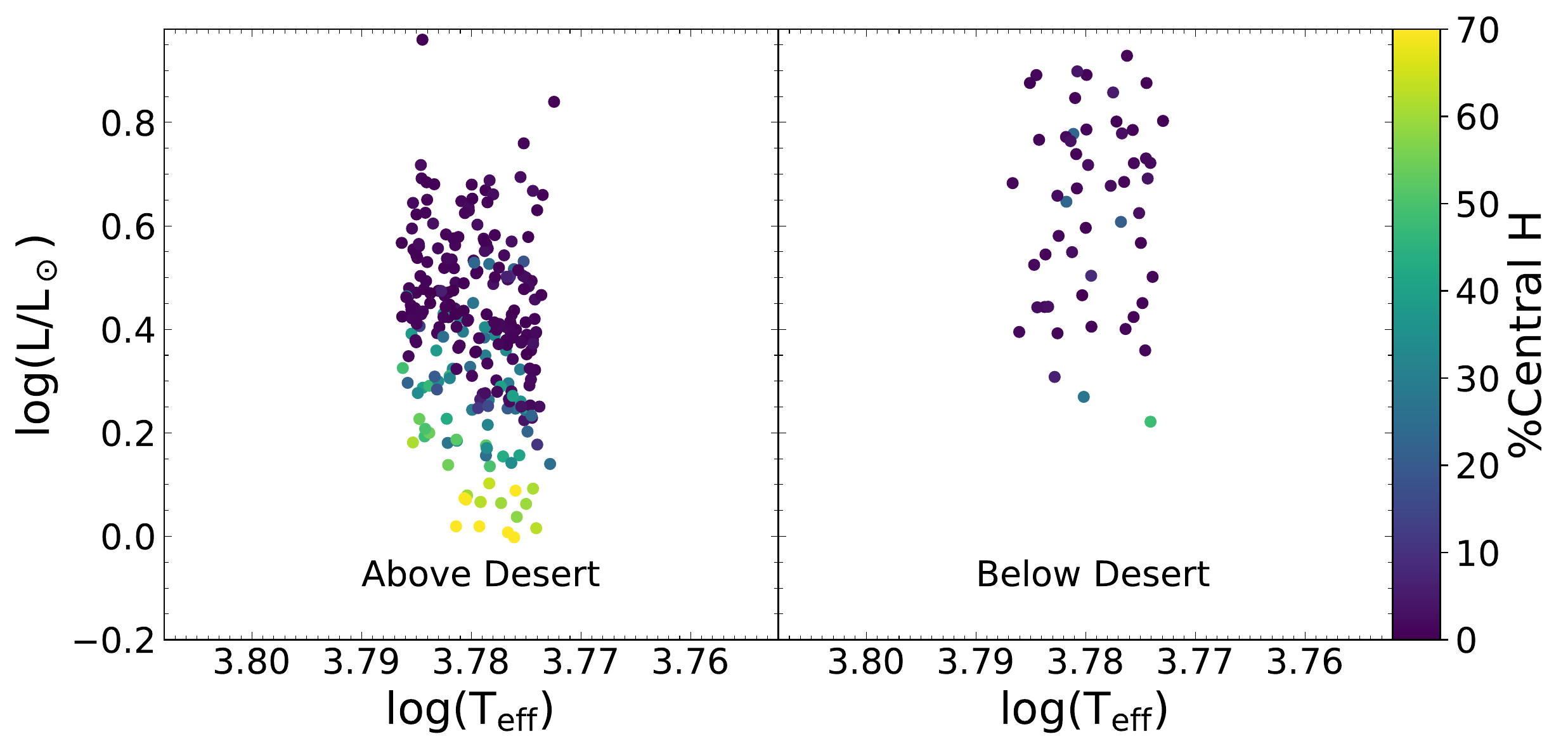}
\caption{HR diagram of stars above (left panel) and below (right panel) the Li 
desert. Colors indicate the percentage of hydrogen in the core of stars, which 
show that those below the desert tend to have less core hydrogen (lighter colors) and thus that 
are more evolved than those stars located above the desert. Stars below the 
desert are also more luminous, indicating that there could be a mass effect 
that we should consider.}
\label{HRcentral}
\end{center}
\end{figure*}

\subsection{Characterizing the Li desert: Mass, metallicity, and evolutionary stage}

Using the extended catalog presented in this work, we identify stars that populate the lower and upper regions of the Li desert, focusing on stars in the specific temperature range of the desert in this subsection (See Appendices \ref{overview} and \ref{charac}  for characterization of the entire sample and the zone around the Li desert, respectively). We also analyze their properties to find differences between stars with high and low Li abundances.

In Figure \ref{Hcentral}, we show from left to right the distributions for mass, metallicity, age, and core hydrogen abundance, representing the exact evolutionary phase, for stars above and below the desert. Top and bottom panels show the frequency and cumulative distribution, respectively.

To obtain the core hydrogen abundance we used stellar models computed with the Yale Rotating 
Evolutionary Code \citep[YREC,][]{Demarque2008} for the mass and metallicity of 
each of the stars. Then, we identify the time step in the model where the 
effective temperature and luminosity match the effective temperature and 
absolute magnitude measured for that star (after applying the corresponding 
corrections). We use both these parameters, instead of age, to locate the evolutionary time step we need, simply because ages obtained by isochrone fitting can be very uncertain. 
The final model selected provides information about the composition of the stellar core, which is used to produce the right panel of Figure \ref{Hcentral}.

The mass distribution in Figure \ref{Hcentral} shows that the stars below the desert are overall more massive than those above. The stars above the desert have masses from $0.8$ to $1.35\,\mathrm{M_\odot}$, while the lower part of the Li desert is populated by slightly more massive stars, with masses from $\sim1.1$ to $1.55\,\mathrm{M_\odot}$. For a detailed characterization of the mass and metallicity of stars surrounding the Li desert (and not only in a limited $\mathrm{T_{eff}}$ range), see Appendix \ref{charac}.

The distribution of metallicity is very similar for both groups of stars. The age distribution is also similar for both populations, although in this last case, stars above the desert are slightly older. Given the complications associated to discussing in terms of age and the large uncertainties they have, whenever possible, we frame our analysis and discussion around stellar mass and evolutionary stage (i.e., hydrogen core fraction), thus avoiding concepts such as young and old. 

The rightmost panel of Figure \ref{Hcentral} shows the distribution of hydrogen in the core of the star. This is used as a more precise indicator of the 
evolutionary stage for each of the stars in the $\mathrm{T_{eff}}$ range of the desert, where more evolved stars have a lower percentage of central hydrogen. As Li is highly dependent on stellar parameters such as temperature, age, metallicity, and mass, the use of any of those parameters to make conclusions on what is producing the Li desert could introduce biases in the analysis, so we have instead decided to use a better indicator of the evolutionary stage of the star, regardless of any of those other stellar properties: the hydrogen fraction in the core.
 
The distributions of core hydrogen content for stars above and below the desert presented in Figure \ref{Hcentral} also appear to be different, with stars in the lower part of the Li desert concentrating towards lower core hydrogen fractions being more evolved.
The main difference between both distributions is given by the fraction of stars with almost zero core hydrogen fraction, but the complete lack of stars below the desert with more than $30\,\%$ core hydrogen separates the two distributions even more.

The HR diagram in Figure \ref{HRcentral}, color-coded according to core hydrogen percentage, emphasizes the same conclusion obtained in the right panel of Figure \ref{Hcentral} regarding evolutionary stage of stars above and below the desert. In this figure we also see how mass is an important factor to consider, as stars above and below the desert have different luminosities, which can be attributed to a mass effect. Additionally, more massive stars have shorter MS lifetimes that can locate stars closer to evolved stages of evolution.

 \textit{We conclude that mass is an important property shaping the Li desert, but also that stars below the desert have usually already left the MS, having evolutionary statuses that are  different from stars with higher Li abundances in that $\mathrm{T_{eff}}$ range.}

\subsection{The origin of the Li Desert}

At the effective temperature of the Li desert, we expect stars to have shallow convective envelopes, and thus, higher Li abundances. Consequently, the stars that seem to have an ``abnormal'' Li abundance are those below the desert. Below, we explore two possible origins for stars in the lower side of the Li desert. One possibility that has been suggested in the literature (e.g., R12) is that there is a depletion mechanism acting inside stars when they evolve, close to the subgiant phase. This depletion mechanism would cause stars in the upper side of the desert to decrease their Li considerably or even completely, moving them in the $\mathrm{A(Li)}$-$\mathrm{T_{eff}}$ plane below the desert. The second possibility we address is that these subgiants on the lower side of the desert are simply evolving from MS stars that originally had less Li; for example, those from the Li dip. Below, we argue that this is indeed the most likely explanation.

To find the origin of the Li desert and where the evolved stars below the desert come from, we disentangle the effects of stellar parameters, especially mass and metallicity, from the evolutionary stage.

\subsubsection{Do stars below the desert evolve from those above it?}

First, we explore the possibility that stars below the desert have evolved from stars above it rapidly decreasing their Li abundances as they get to the subgiant phase. As we see throughout this section, we conclude that the distribution of stellar parameters of stars with high and low Li are different, and the progenitors of stars with low Li abundances (below the desert), are not located in this same effective temperature range. Thus, stars below the Li desert do not evolve from those above.

\begin{figure}[!hbt]
\begin{center}
\includegraphics[width=0.5\textwidth]{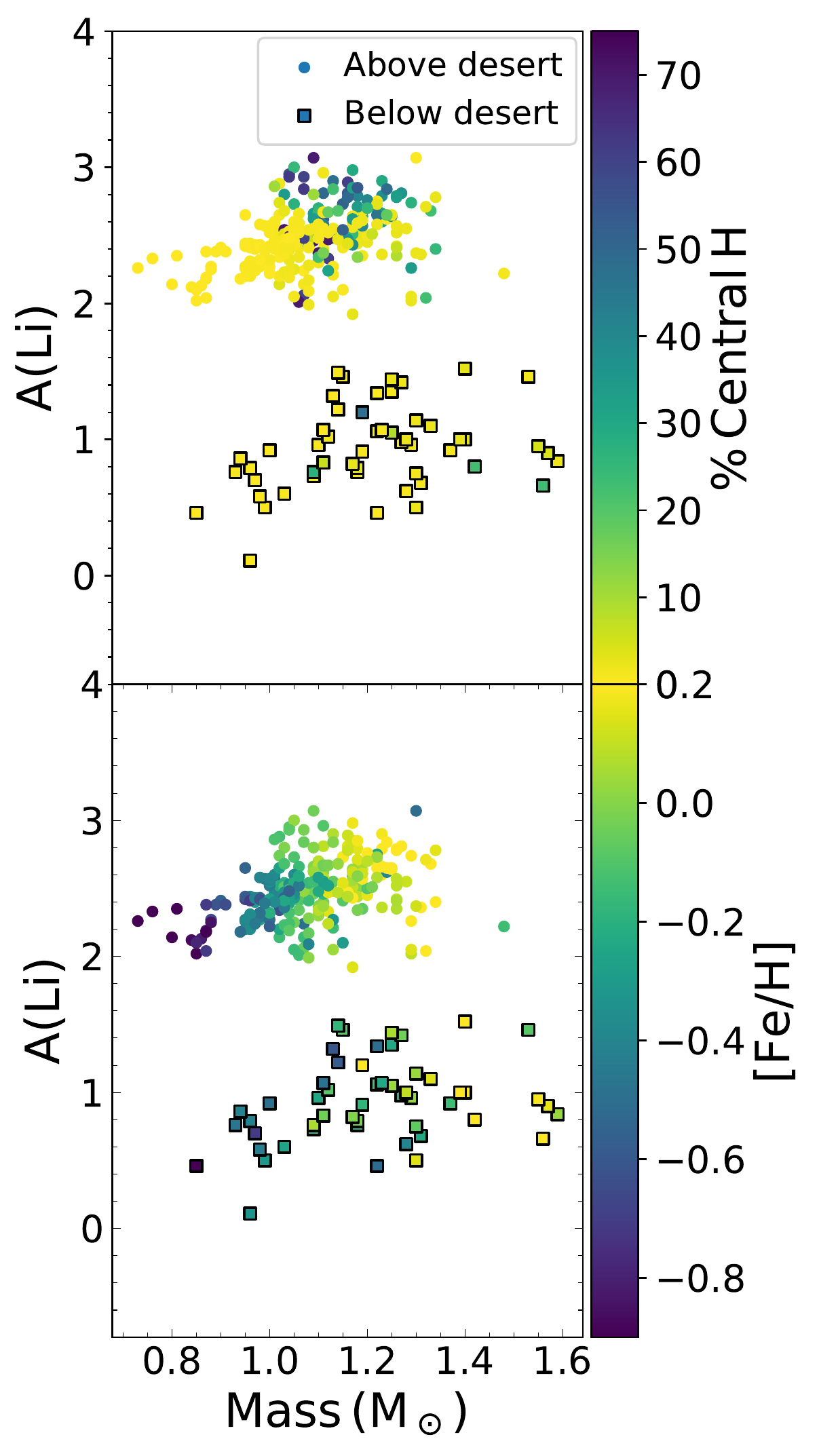}
\caption{Lithium as a function of mass for stars in the effective temperature range 
of the Li desert, between $5950$ and $6100$ K, color coded by their percentage 
of central hydrogen (top) and metallicity (bottom). Different symbols are used 
to depict stars above (circles) and below (squares) the Li desert.}
\label{mass_age}
\end{center}
\end{figure}

We analyze the lithium as a function of mass for all stars in the temperature range 
between $5950$ and $6100$ K in Figure \ref{mass_age}, to understand how the 
different distributions of mass, metallicity, and evolutionary stage affect the Li abundance. 
In the top panel, stars are color-coded by the hydrogen percentage in their core. Also the different 
symbols are to identify stars above (circles) and below (squares) the Li 
desert. The top panel emphasizes that stars below the desert are mostly evolved (lighter colors). In the mass range between $1.0$ and $1.3\,\mathrm{M_\odot}$, we can find stars both above and below the desert, indicating that there must be something else besides mass contributing to their morphology. Even in this common mass range, stars below the desert have, overall, a lower percentage of core hydrogen.

\begin{figure*}[!hbt]
\begin{center}
\includegraphics[width=0.55\textwidth]{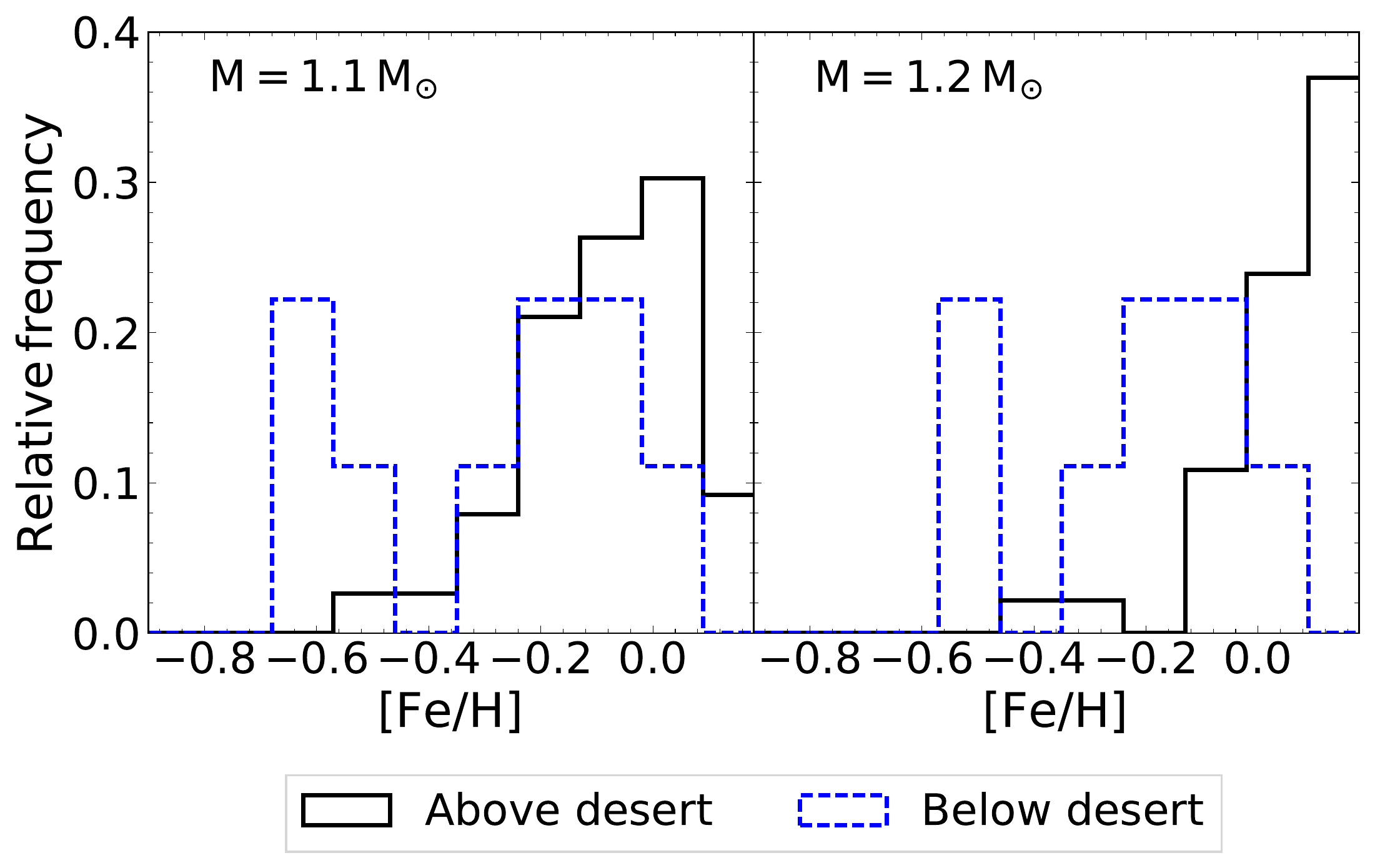}
\caption{Distribution of metallicity for stars above and below the desert in a specific range of masses. The left panel includes only stars with $1.10\pm0.05\,\mathrm{M_\odot}$, and the right panel shows stars with $1.20\pm0.05\,\mathrm{M_\odot}$. We can see that stars above the desert are consistently more metal rich than those below at the same mass. As such, these two groups of stars are different in their fundamental properties, and no further attempts can be made to find the origin of the subgiants with low Li in the stars above the desert, with high Li abundances.}
\label{mass1112}
\end{center}
\end{figure*}
 
To find out whether or not stars below the Li desert evolve from those above, that is, whether or not an additional depletion process is acting on stars as they evolve to the subgiant branch, we want to compare the stars below the desert with MS stars of similar mass and metallicity, which are expected to follow similar evolutionary processes (as long as there is no extra-mixing or other involved additional parameters, such as initial rotation rate).
However, as we can see in Figure \ref{mass_age}, we do not find MS stars in the entire mass range covered by stars above and below the Li desert. Both the most massive stars (M$>1.3\,\mathrm{M_\odot}$) 
and the least massive (M$<1.0\,\mathrm{M_\odot}$) stars forming the desert are subgiants, meaning that we can find almost no MS stars in these mass ranges at the temperature of the desert. This is explained by how stars that belong to the Li desert are selected, only based on their effective temperatures. Stars of higher masses (from $\sim1.3\,\mathrm{M_\odot}$ and above) during the MS have higher temperatures than those defined for the desert. The same is the case for the lower mass range, where, depending on the metallicity, the stars in the MS can be colder than $5950$ K, not falling in the region of the Li desert.

 \begin{figure*}[!hbt]
\begin{center}
\includegraphics[width=\textwidth]{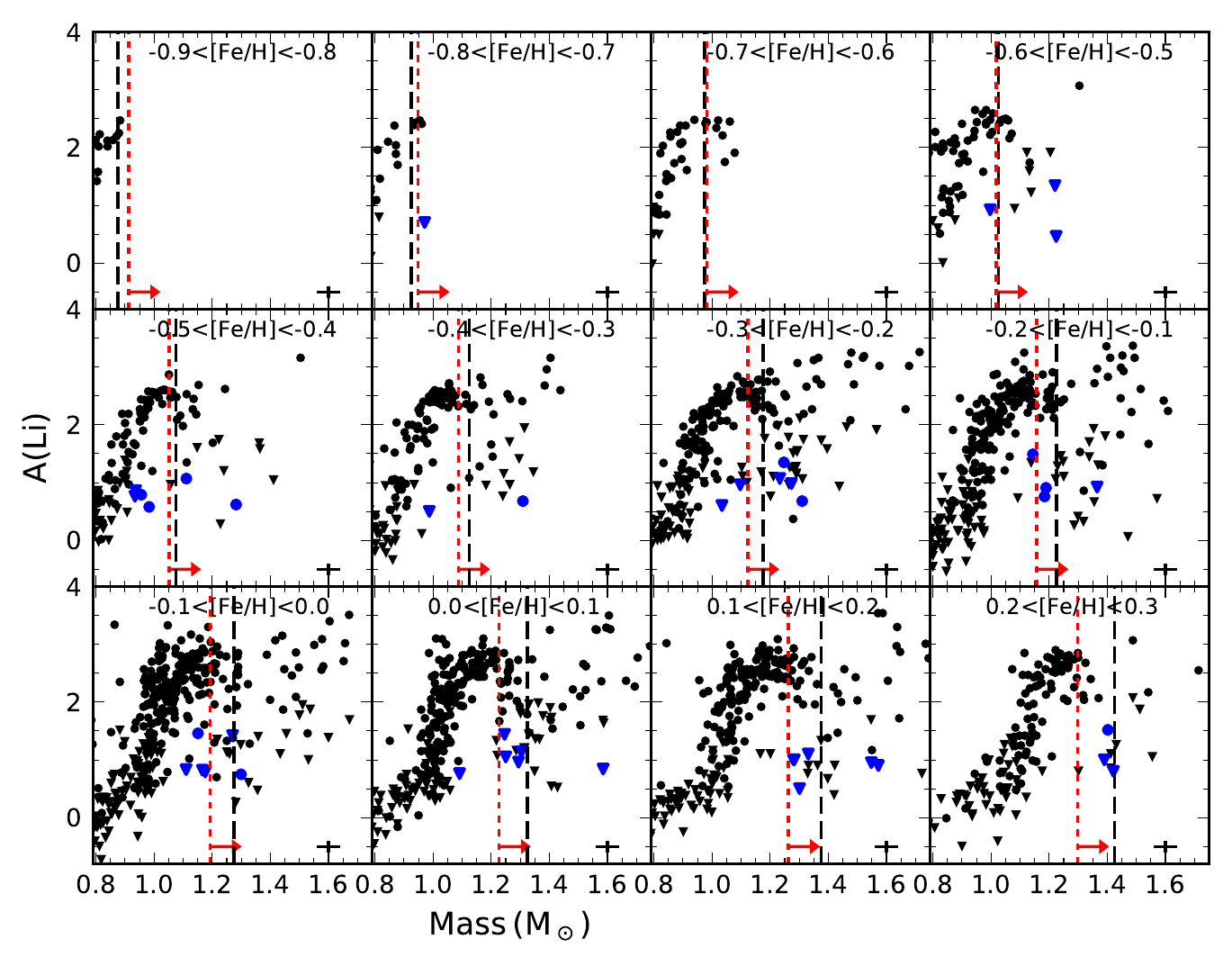}
\caption{Li as a function of mass for all stars in the sample, in different 
metallicity bins, where the blue symbols represent stars located below the Li desert. Downward triangles are Li upper limits. Each panel shows the location of the center of the Li dip (dashed black line) and the cooler side (red dotted line) based on a fit obtained by the analysis of clusters by \citet{Balachandran1995} and \citet{Cummings2012}, respectively. As in previous figures, circles are Li determinations and triangles are upper limits.}
\label{dip}
\end{center}
\end{figure*}

The bottom panel of Figure \ref{mass_age} shows the same Li abundance-mass relation as the top panel, now color-coded according to metallicity. We see that stars above and below the Li desert have different properties. Below the desert (squares, low Li) we do not find stars with [Fe/H]<$-0.45$, limiting the range of stellar parameters where we find stars above and below the desert. For a fixed mass, stars below the desert are consistently more metal poor (darker colors in the bottom panel of Figure \ref{mass_age}) than those above it. We illustrate this more directly in Figure \ref{mass1112}, where we focus on two different small mass ranges in the left $(1.10\pm0.05\,\mathrm{M_\odot}$, 57 stars in total) and right panels $(1.20\pm0.05\,\mathrm{M_\odot}$, 87 stars in total), and see how the distribution of metallicity looks for stars above and below the desert. In both cases, stars above the desert are more metal rich. 
Having the same mass, but differing in metallicity implies that these stars follow different evolutionary tracks and can experience different degrees of Li depletion, and thus, they should not be compared.

Therefore, given that in the effective temperature range of the desert stars in the MS have different stellar parameters (mass and metallicity) than slightly evolved subgiants, they could also have different initial Li abundances, and it is incorrect to compare them directly. As such, stars below the desert have not evolved from those above. The correct comparison to be made is between stars below the desert and their actual progenitors, that is, stars that in the MS have similar masses and metallicities and that may or may not be in the effective temperature range of the desert. Stars selected by this blunt cut in effective temperature do not provide the necessary information to make any robust conclusions.

\subsubsection{Do stars below the desert evolve from the Li dip?}

Now we explore the locations of the progenitors of the stars below the desert, and if they already have less Li on the MS. To do this, we include the full catalog, without selecting stars by their effective temperature. This highlights the relevance of our full homogeneous sample.
We conclude in this section that stars below the desert are consistent with having evolved from the Li dip.

Figure \ref{dip} shows $\mathrm{A(Li)}$ as a function of mass for all the stars in our sample in different metallicity bins. As the Li dip was originally identified in open clusters, with restricted metallicities, we replicate these conditions with field stars, creating a situation analogous to that found in clusters, by creating clusters of field stars in each panel. For clusters, the Li dip in this diagram is shown as an overall decrease in the Li abundance at a given mass. The Li dip is not as easily identified in the field as in clusters, and although it is not seen as a clear dip in Li abundance, we can see an increase in the amount of Li upper limits at a mass range specific for each metallicity bin (i.e., in each panel). 

In general, for a given metallicity bin in Figure \ref{dip} we see that less massive stars have lower Li abundances than their higher-mass counterparts. 
However, we can identify a certain mass in each bin where the Li abundance decreases (e.g., for $-0.4<$[Fe/H]$<-0.3$, between masses $\sim1.1$ and $1.3\,\mathrm{M_{\odot}}$). This decrease in Li abundance is the Li dip, and the mass at which it is found is consistent with that reported for 
different metallicity bins in other works in clusters \citep{Balachandran1995,  AnthonyTwarog2009, Cummings2012, Francois2013}, and less clearly, in the field \citep[e.g.,][]{LambertReddy2004, DelgadoMena2015}. We show the approximate position in mass of the Li dip in Figure \ref{dip} by two vertical lines. These are calculated in each panel by taking the middle metallicity of each bin. The black dashed line is the center of the dip, based on the fit by \citet{Balachandran1995}, where $\mathrm{M/M_\odot}=1.3+0.5$[Fe/H]. The red dotted line is the cool side of the dip based on the linear fit by \citet{Cummings2012}, where $\mathrm{M/M_\odot}=1.21+0.35$[Fe/H]. We also include an arrow pointing to the direction of the Li dip. In this last work they provide the position of the cool side of the dip, because they study older clusters where stars in the hotter side of the dip have already evolved. As such, the hot side (and, consequently, the center) is difficult to identify.

The correlation of increasing mass of the dip with higher metallicity that has previously been found in the literature is also present. Once again, we stress that the Li dip is less prominent in our sample of field stars than in clusters. Here, it is seen as a scatter in the $\mathrm{A(Li)}$ at the expected mass of the dip, instead of a sudden decrease in abundance for all stars.

This representation of the Li dip, with stars in the lower side of the Li desert shown in blue in Figure \ref{dip}, allows us to compare the position of these stars with others of similar parameters in each panel. We see that the stars below the desert are not unusual and are generally located among other stars of the sample. Thus, \textit{other stars with the same mass and metallicity also show low Li abundances}. This indicates that stars located below the Li desert are not unusual and do not require extra-mixing or enhanced depletion to be explained. They are consistent with having evolved from stars already present in the sample.

Moreover, the position of the stars below the desert is, in most cases, consistent with the Li dip. Thus, these stars are consistent with having evolved from the Li dip and that possibility cannot be discarded. Some exceptions can, however, be identified. In the panel $-0.5<$[Fe/H]$<-0.4$, four of the stars located below the Li desert have very low masses, inconsistent with the dip. Other stars at similar masses tend to have higher Li abundances. The Li abundance of these four stars could be consistent with that of lower-mass objects of $\sim 0.9\,\mathrm{M_\odot}$, within the uncertainty of the estimated masses. We suggest, then, that the $\mathrm{A(Li)}$ of these four stars does not need further mixing mechanisms to be explained. In contrast, stars like the dwarf HIP22589 with near solar metallicity, HIP50805 with [Fe/H]$=-0.26$, or HIP49988 with [Fe/H]$=-0.5,$ seem to be very different to other stars surrounding them in these diagrams, and are depleted when compared to stars with similar mass, metallicity, and even age. HIP49988 is defined as an s-process enriched star \citep{Zenoviene2015} and HIP50805 is a CH subgiant \citep{LuckBond1982}, also enriched in s-process elements. Therefore, both are chemically peculiar stars thought to be produced by mass transfer from an evolved companion. Furthermore, the mass-transfer episode can alter the chemical abundances of the stars, explaining their Li content. HIP22589 is one of the few stars with low Li abundance in the temperature range of the desert classified as a MS star considering the percentage of central hydrogen, so it is still somewhat peculiar.

Other stars below the Li desert are surrounded by stars with the same masses and metallicities that also show low Li abundances, so they could be explained by stellar evolution from these already depleted progenitors.

As we have concluded that stars above and below the desert are intrinsically different in terms of their initial physical properties, the position of the exact Li desert is of little importance, because the desert is simply an artifact, and its definition is artificial. If the stars below the desert are actually evolving from the Li dip, then their position in the $\mathrm{A(Li)}$-$\mathrm{T_{eff}}$ is coincidental. These stars would not be peculiar and we can distinguish them only because the combination of atmospheric parameters, and the decrease of their $\mathrm{T_{eff}}$ as they cross this diagram while evolving, locates them in a particular region of parameter space, where otherwise, only stars with high Li abundances would be found.

\section{Discussion} \label{disc}

Our main conclusion is that most of the low-Li-abundance stars in the 
temperature range $5950$ to $6100$ K are consistent with having evolved from the Li dip. Even if this 
temperature range is not often studied, some other works have previously 
identified stars that are supposed to be evolving from the MS Li dip, but this work makes the connection quantitatively sound by demonstrating that low-Li stars tend to be more evolved, in the subgiant branch or close to it.

The sample of Li abundances presented by \citet{Balachandran1990} is subdivided 
into groups based on the rotation periods of stars. In this work, no Li desert 
is defined in the $\mathrm{A(Li)}$-$\mathrm{T_{eff}}$ plane, since upper limits 
were defined at a much higher Li abundance and covered that entire region. In that work, some slowly rotating Li depleted stars are found. When placed on an HR diagram, they are surrounded by stars with larger Li abundances. As the Li-depleted stars have a similar mass to 
the Li dip, \citet{Balachandran1990} suggest that they are actually field 
analogs to that feature, i.e., they have evolved from MS stars with 
temperatures between $6500$ and $6800$ K. Although this might be the case,  being hotter,
these stars in particular are not found in the temperature range of the Li 
desert,  so they do not correspond to the subgiants studied in this 
work.

In \citet{Lambert1991} the desert is not clear, but they reported that stars 
with temperatures higher than $5900$ K tend to have a bimodal distribution in 
the $\mathrm{A(Li)}$-$\mathrm{T_{eff}}$ plane, as one group has high Li
determinations, and the other only had upper limits. They suggest that the low-Li group could have evolved from stars with ZAMS $\mathrm{T_{eff}}=6600$ K, 
corresponding to the Li dip. Once again, the temperature range of the stars 
considered is higher than that studied here.

\citet{Chen2001} have also studied the Li desert, where they can clearly 
identify a gap for $\mathrm{T_{eff}}>5900$ K. They mention that a rapid 
mechanism of Li depletion must be acting inside those Li-poor stars. Given that 
they are located in a very narrow mass range, and that they follow a 
mass-metallicity correlation similar to that suggested for Li dip stars 
\citep{Balachandran1995}, \citet{Chen2001} conclude that the Li-poor stars are evolved from 
the Li dip. In this work, we found that all stars in the sample follow a similar mass-metallicity relation regardless of their Li abundance (R12), and therefore it would be impossible to conclude that Li-poor stars below the desert are evolved from the dip based on this evidence alone. However, from a completely different study of these stars, 
we arrive at the same conclusion.

An important conclusion behind the origin of the Li desert is that field stars 
that have evolved or belong to the Li dip cannot be identified exclusively from 
their effective temperatures, and the important parameter in determining their 
Li depletion is the mass \citep{Chen2001}. As we can see, their metallicity 
also plays an important role.

\section{Summary} \label{summ}
The Li abundance of stars allows us to study several different phenomena related 
to stellar interiors, cosmology, and even planets. Thus, it is necessary to 
have a large amount of stars to study these problems adequately, but also, it is necessary that both the atmospheric parameters and Li 
abundances are internally homogeneous.

In this work, we report measurements of surface Li abundance and 
atmospheric parameters of 227 stars, and compile a homogeneous catalog of Li 
abundances based on the previous catalog by R12 but adding the large datasets 
presented in recent works, such as \citet{DelgadoMena2014, DelgadoMena2015} and 
\citet{Gonzalez2014, Gonzalez2015}. All stars considered have Hipparcos 
paralaxes that allow an internally consistent determination of masses and ages 
in our sample of 2318 stars, including dwarfs and subgiants.

The region of mid-F stars, where the Li dip is found, and the region of G 
stars, where we find objects with similar characteristics to our Sun, have been 
extensively studied, but the effective temperature range between has not 
received as much attention (between $5950$ and $6100$ K). Nevertheless, 
previous studies in this temperature regime have found a peculiar behavior in 
the Li abundance of stars, as they tend to separate discontinuously into high 
and low Li abundances. The region in the $\mathrm{A(Li)}$-$\mathrm{T_{eff}}$ 
where no stars are found is the Li desert. With our extended catalog, we identify the region of the Li desert that separates stars that have high and low Li abundances. Given the increased sample size, this feature is probably not produced by a selection effect. 
In the region of the Li desert, given the effective temperatures, canonical stellar evolution predicts stars with high Li abundances, and therefore it is 
those stars below the Li desert that are the unusual objects requiring explanation.

It has been suggested that the low-Li-abundance side may be produced by stars 
that have evolved from the Li dip, but R12 recently questioned this claim and 
proposed a different alternative, where there is a depletion mechanism acting 
inside the MS as they evolve, or subgiant stars that are producing the low-Li-abundance side of the 
desert.

We focus on the temperature range of the Li desert to study the origin of the 
low-Li-abundance stars. Most of the observed stars presented in this catalog 
are located in this region (between $5950$ K and $6100$ K).
With this catalog, we analyze the suggestion by  R12 that the Li desert may be related to the presence of planets around stars, given that planet hosts in their sample were only found in the region of high Li abundance. Although we cannot discard that the presence of planets may affect the Li abundance of stars, by increasing the sample size in the region of the Li desert we have found that this effect seems to be coincidental.

Using the extended sample combined with stellar evolutionary models, we conclude that the majority of the stars in 
the lower side of the desert are more evolved than those with high Li.

Features and correlations of stars around the Li desert with mass, age and 
metallicity can be summarized as follows:  
Stars of different masses populate the upper and lower parts of the Li desert, where the latter includes more massive stars. Although both populations above and below the desert share the same range in metallicities, at a fixed mass, those below the desert are consistently more metal poor than stars with high Li abundance. There also seems to be a slight age difference between stars above and below the desert, with those 
below being younger than those above.

We have concluded that stars above and below the desert have intrinsically different parameters, therefore we conclude that stars with low Li abundances in the effective temperature range of the desert do not evolve from those with higher abundances by a very rapid depletion mechanism acting close to the subgiant. In contrast, we suggest that making a cut in effective temperature leaves out the progenitor sample of the stars in the lower side of the desert.

As such, we extend the analyzed sample to include all FGK stars in our catalog to find the origin of the Li desert. Although the feature of the Li dip is less evident in field stars, by grouping stars in the field analogous to clusters (with fixed metallicities but different masses), we can identify the point in mass where the spread in Li abundance increases. Thus, we recover stars from the Li dip, and see how the mass of these low-Li-abundance stars changes with 
metallicity, a behavior already identified by \citet{Balachandran1995}. 

Stars of similar mass and metallicity should follow similar evolutionary 
tracks. Consequently, by removing effects from 
stellar parameters, we can constrain the origin of the subgiants in the low side of the Li 
desert and confirm that all of them are surrounded by stars of similar stellar parameters. Consequently, they do not seem to be special or require further explanation. Moreover, most of the stars below the Li desert have a combination of mass and metallicity that places them in the same location as stars in the Li dip. As such, these stars with low Li below the desert are consistent with being evolved from Li dip stars.

What produces the depletion in stars of 
the Li dip remains to be elucidated. Candidates for the physical origin behind this dip are rotationally induced mixing \citep{Pinsonneault1990, DeliyannisPinsonneault1997}, mass loss \citep{Schramm1990}, and diffusion \citep{RicherMichaud1993}, but no consensus has yet been reached.

Although we have considerably increased the sample size with this work, with the \textit{Gaia} spacecraft and the new parallaxes
it will provide, along with abundance measurements from high-resolution 
spectrographs, we will be able to study this problem with an extensive and 
homogeneous catalog, and eventually find a physical reason 
for the depletion.

\begin{acknowledgements}
We thank the referee Patrick de Laverny for comments and suggestion that helped to improve the quality of this work. We thank Prof. D. L. Lambert for supporting partially this work as well as our visits to McDonald Observatory.
This research has made use of: TOPCAT \citep{topcat}, numpy, matplotlib, the 
VizieR catalogue access tool (CDS, Strasbourg, France), and The Extrasolar 
Planets Encyclopaedia. Support for C.A-G was partially provided by CONICYT-PCHA 
Doctorado Nacional 2013-21130353. C.A-G and JC acknowledge support from the 
Chilean Ministry for the Economy, Development, and Tourism's Programa 
Iniciativa Cient\'ifica Milenio, through grant IC120009 awarded to the 
Millenium Institute of Astrophysics (MAS) and from PFB-06 Centro de Astronomia 
y Tecnologias Afines.
\end{acknowledgements}

\bibliographystyle{aa}
\bibliography{Lidesert}{}

\begin{appendix} %First appendix
\section{An overview of the catalog}\label{overview} 

\begin{figure}
\begin{center}
\includegraphics[width=0.5\textwidth]{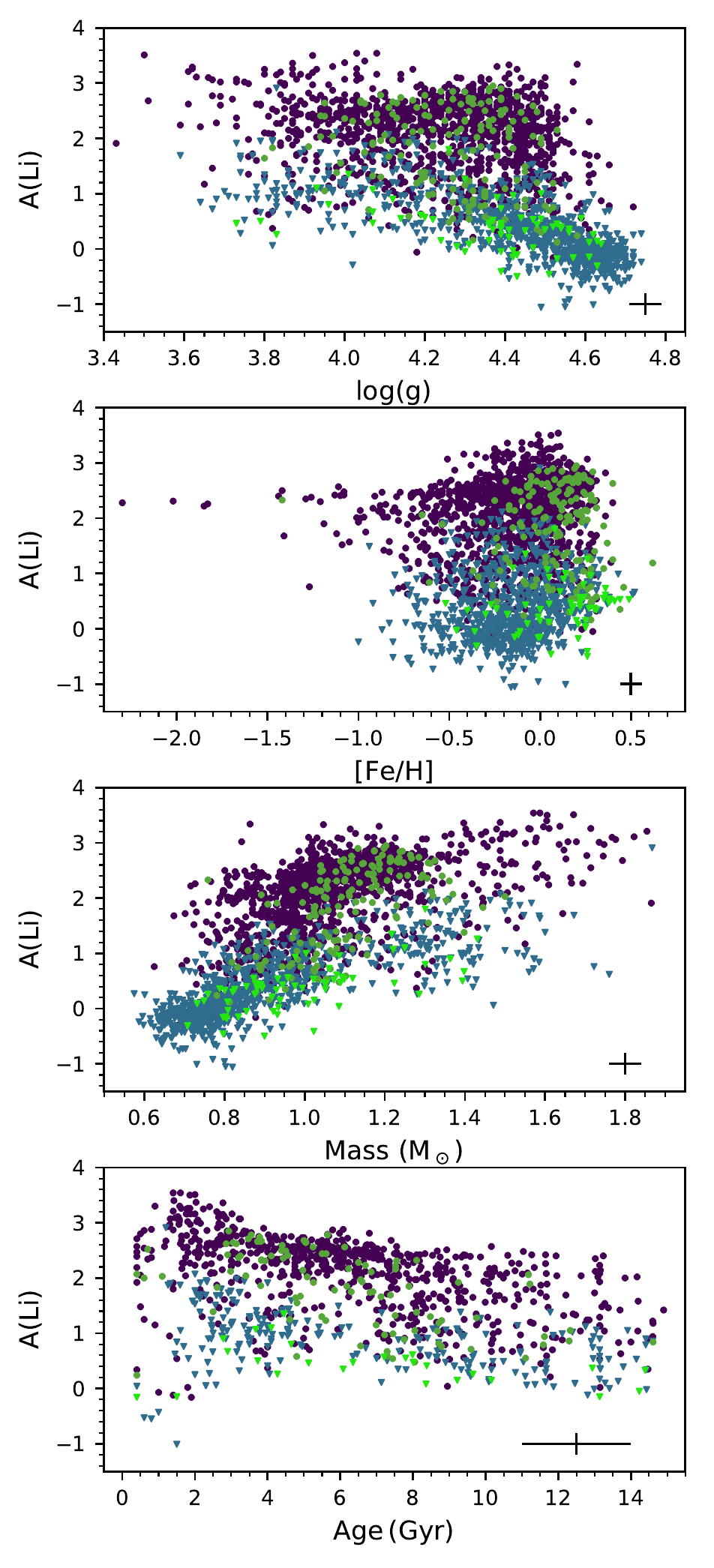}
\caption{Li abundance of stars in the final catalog, as a function of $\log g$, [Fe/H], mass, and age (from top to bottom panel). Circles 
(triangles) represent Li abundance determinations (upper limits), and the 
sources are also color-coded according to their planet-hosting status. Green 
points are planet hosts, while blue points are the comparison sample, including 
non-hosts and stars with unknown status. The bottom panel only 
shows stars with reliable ages.}
\label{AllvsAll}
\end{center}
\end{figure}

Figure \ref{AllvsAll} presents the Li abundance of stars in the compiled 
catalog, and their distribution with $\log g$, [Fe/H], mass, and age. 
In the bottom panel, we only use stars with reliable ages where age/(error age)$>3$, which correspond to 1093 stars or $47\%$ of the total sample. Stars are color-coded based on their planet hosting status (green represents planet hosts) and if Li is detected in the star (darker colors) or only an upper limit could be found.

Figure \ref{AllvsAll} also allows us to identify some features of our sample. 
Regarding metallicity, most of the stars have metallicities near solar, with 
only a few outliers at [Fe/H]$<-1.0$. It can also be seen that planet hosts 
tend to concentrate towards higher values of [Fe/H]. A large number of studies 
have confirmed this correlation \citep[e.g.,][among others]{Gonzalez1997, FischerValenti2005, Bond2006}, although it seems this is only applicable to 
stars hosting giant planets and not those hosting Neptune or Earth-like planets 
\citep{Sousa2008, Buchhave2015}. 

\begin{figure}
\begin{center}
\includegraphics[width=0.5\textwidth]{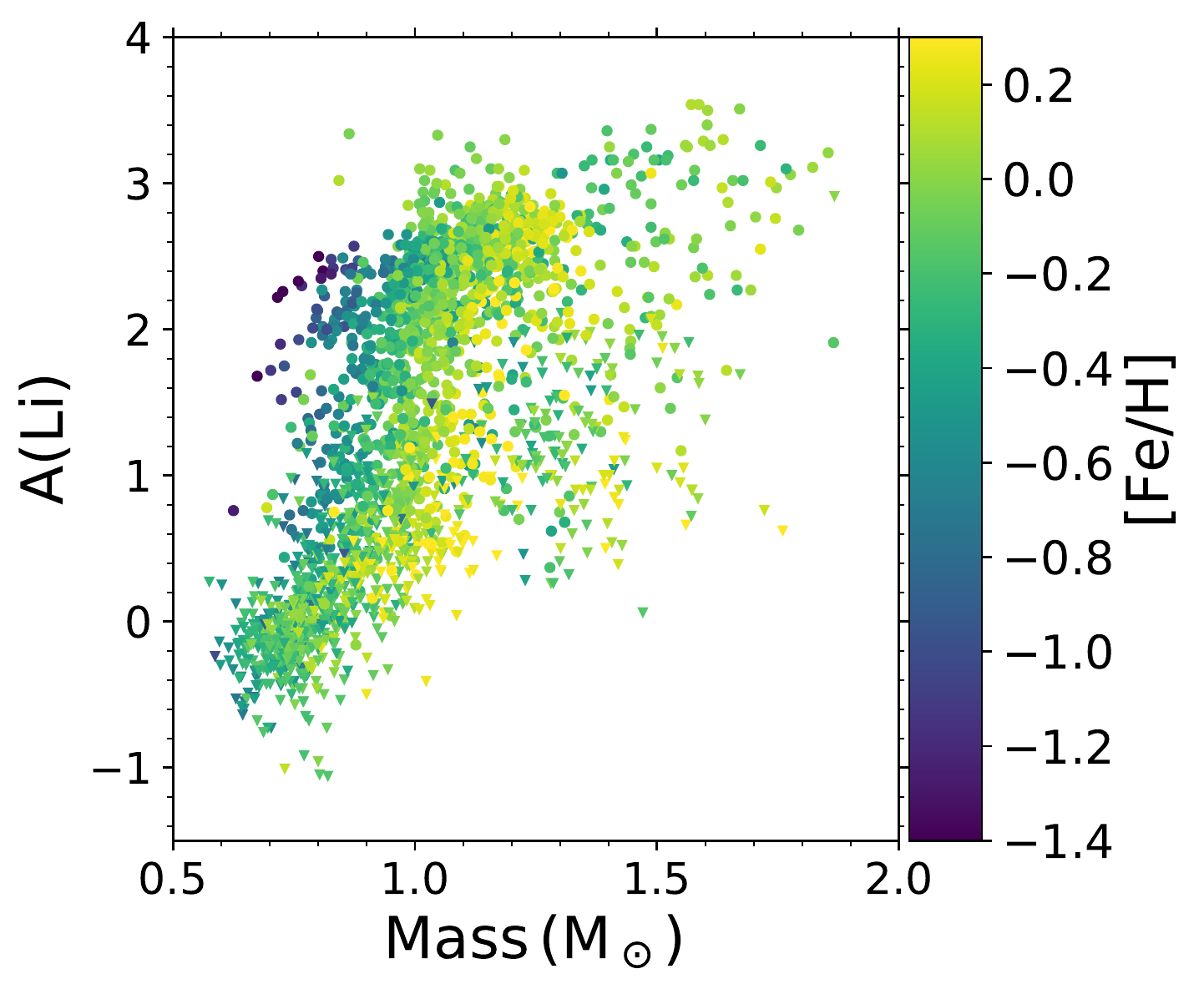}
\caption{Li abundance of stars in the final catalog as a function mass, color-coded according to their metallicity. Circles 
(triangles) represent Li abundance determinations (upper limits).}
\label{AlivsMass}
\end{center}
\end{figure}

The relation of Li abundance with mass shows a positive correlation, although with some dispersion. The trend can be explained if we consider that our sample consists of dwarfs and slightly evolved subgiants. For this type of star, higher-mass objects have a shallower surface convection zone and thus experience less Li dilution. Moreover, the temperature at the base of the convective zone, which determines the amount of Li depletion, 
also depends on metallicity. Thus, it is not unexpected to find a spread in lithium at each given mass, and even when other stellar parameters as metallicity and age are fixed, there is still a large spread in Li abundances. We can confirm this with Figure \ref{AlivsMass}, showing the same third panel from Figure \ref{AllvsAll} but with stars now color-coded by their metallicity. Finally, in the bottom panel, Li shows a large spread 
at different ages, but there is still a decreasing correlation in the upper envelope, especially at the younger end.

\begin{figure}
\begin{center}
\includegraphics[width=0.5\textwidth]{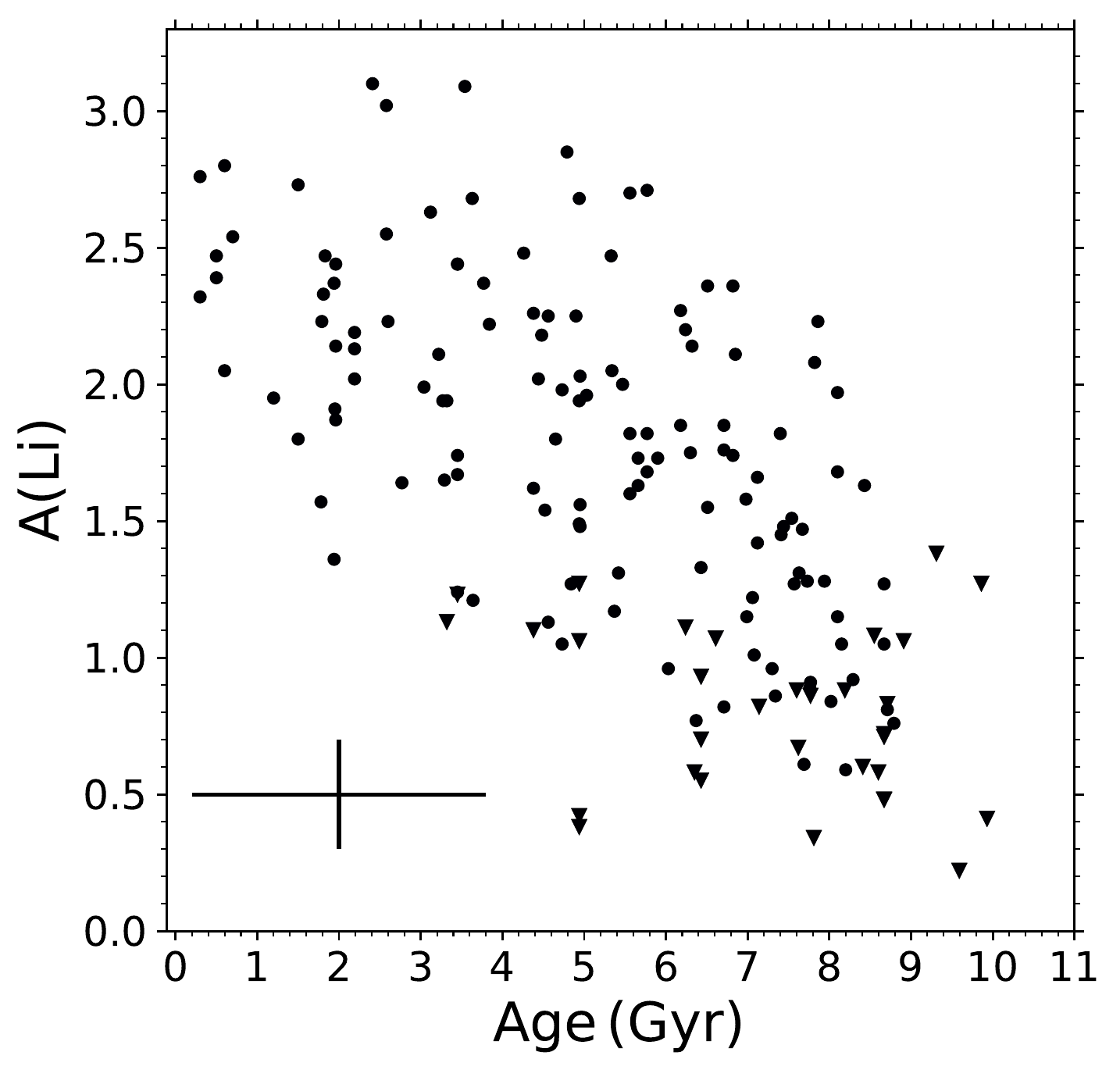}
\caption{Li abundance as a function of age for solar twins in our sample, selected by ranges in effective temperature, $\log g$, metallicity and mass ($T_{\mathrm{eff}}=T_{\mathrm{eff},\odot}\pm90$ K, $\log g=\log g_\odot\pm0.2$, [Fe/H]=[Fe/H]${}_{\odot}\pm0.11$ and M=M${}_\odot\pm0.05$). The points are Li detections and the triangles represent upper limits. We see how Li seems to decrease with age very strongly, but the correlation shows considerable scatter which may be caused by measurement errors.}
\label{Twins}
\end{center}
\end{figure}

If we only select solar twins by restricting ranges of effective temperature, 
$\log g$, metallicity and mass ($T_{\mathrm{eff}}=T_{   
\mathrm{eff},\odot}\pm90$ K, $\log g=\log g_\odot\pm0.2$, 
[Fe/H]=[Fe/H]${}_{\odot}\pm0.11$ and M=M${}_\odot\pm0.05$), we do find a 
decreasing correlation between Li abundance and age but with a large scatter, as can be seen in Figure \ref{Twins}. 
The scatter could be produced by the uncertainties in the determination of stellar parameters, especially by the errors of age determinations. It is also possible, as indicated by \citet{Baumann2010}, that different 
initial rotation rates for stars could produce a spread in Li abundances. When 
comparing their data with models from \citet{CharbonnelTalon2005} they found 
that the different Li abundances at a given age for solar twins could be 
explained by rotation. In the same narrow ranges of metallicities and masses, 
\citet{CarlosNissenMelendez2016} obtain a very tight correlation between Li 
abundance and ages. The difference between this correlation and what we find in Figure \ref{Twins} can be explained by the much smaller estimated errors in stellar parameters found in that work when compared to ours.

From the entire sample, no clear correlation between planet-hosting status and 
Li abundance can be seen, but the correlations found in the literature are 
usually restricted to solar analogs and solar twins. We do not investigate this 
matter further, since finding a difference between the planet hosting sample 
and comparison stars requires an even larger sample of stars, as it is necessary to apply several cuts to 
stellar parameters. Even with this large amount of stars, restrictions in 
every different parameter must be made so that distributions of planet hosts 
and non-hosts are homogeneous and no biases are being included artificially. We 
cannot assure that we will not find any correlation between planet hosting 
status and Li abundance by making other arbitrary cuts to the stellar 
parameters \citep[e.g.,][]{Baumann2010}.

\section{Characterization of stars surrounding the Li desert}\label{charac}
\begin{figure*}[!h]
\begin{center}
\includegraphics[width=\textwidth]{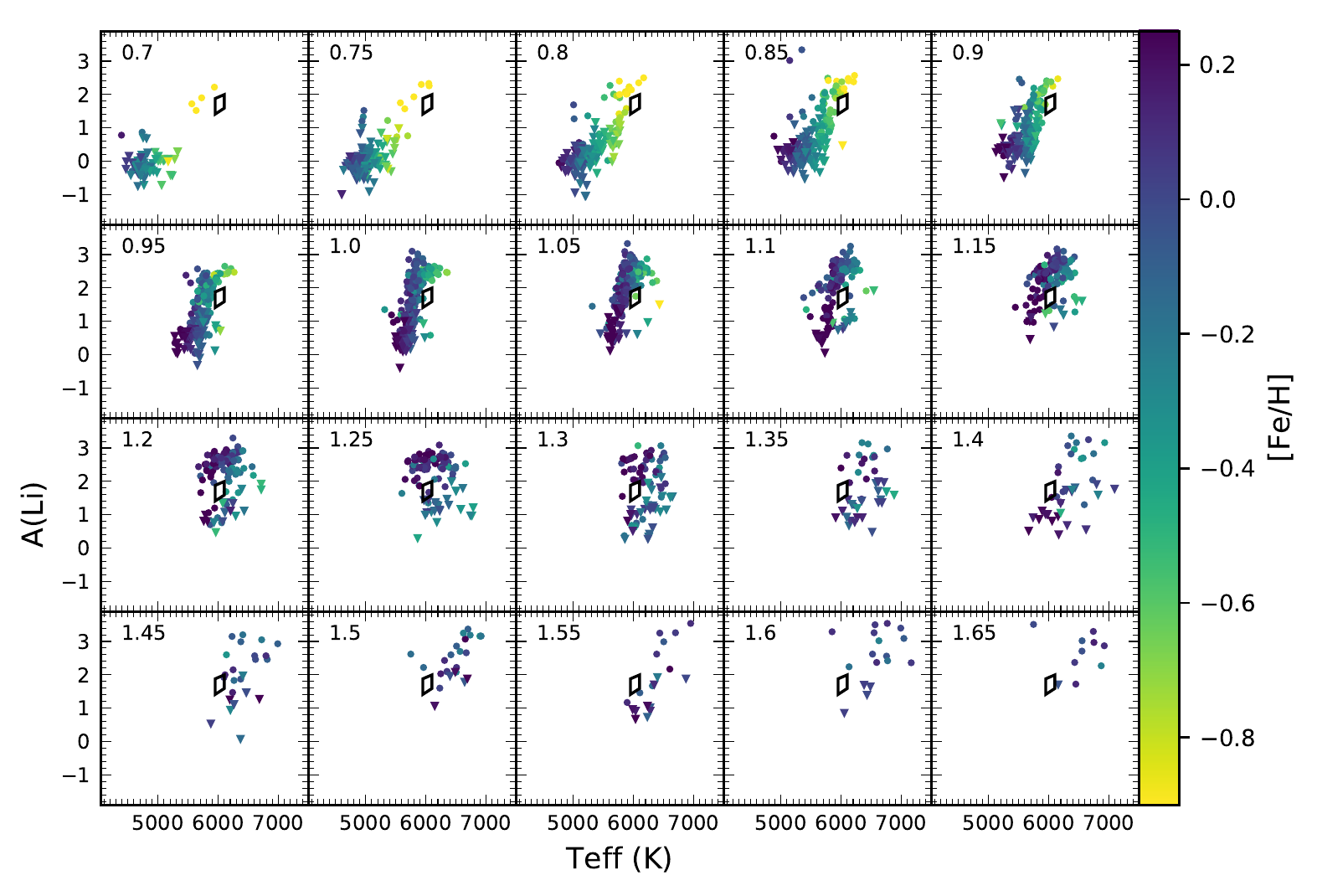}
\caption{Li abundance as a function of $\mathrm{T_{eff}}$ of stars in different 
bins of mass, as indicated in the top left corner of each panel 
(M$\pm0.025\,\mathrm{M_\odot}$). Circles (triangles) represent Li abundance 
determinations (upper limits), and stars are color coded according to their metallicity. 
The polygon shows the approximate location of the Li desert.}
\label{Ap1}
\end{center}
\end{figure*}

The stellar characteristics of stars surrounding the Li desert can help us understand its nature and identify possible trends. We 
explore further the $\mathrm{A(Li)}$-$\mathrm{T_{eff}}$ plane by subdividing 
stars into different mass ranges and using a color gradient to indicate the 
stars' metallicity in Figure \ref{Ap1}. Also, the approximate position of the 
Li desert is indicated by the polygon. 

The known behavior of MS stars is recovered as one examines the series of 
panels of Figure \ref{Ap1}. As the mass of stars increases so does their effective temperature. Also, a very clear metalicity gradient can be seen for stars with lower masses, given because at a fixed mass, more metal-rich stars have larger opacities, which makes them cooler.
This also relates to the surface Li abundance of these stars. The higher opacity of metal-rich stars increases the importance of the surface convection, increasing the size of the convection zone and decreasing the Li abundance. Thus, metal-rich stars tend to be more Li-depleted than their metal-poor counterparts.
These two last trends involving metallicity fade for stars of 
higher mass (M$>1.2\,\mathrm{M_\odot}$). 

From Figure \ref{Ap1}, stars with masses $M<0.8\,\mathrm{M_{\odot}}$ almost do not 
contribute to the formation of the Li desert, only contributing with a few low 
metallicity stars to the upper region. We confirm with this figure, that this side of the desert is mainly 
populated by stars of all metallicities with masses from $0.8$ to 
$1.35\,\mathrm{M_\odot}$, and the lower side with stars of masses from $\sim1.1$ to 
$1.55\,\mathrm{M_\odot}$. As previously noticed by R12, stars in some bins of mass can have high and low Li abundances in the region of the desert. In other words, there is an important effect of mass in the conformation of the Li desert.
\end{appendix}
\end{document}